\documentclass[lettersize,journal]{IEEEtran}
\usepackage{amsmath,amsfonts}
\usepackage{algorithmic}
\usepackage{algorithm}
\usepackage[caption=false,font=normalsize,labelfont=scriptsize,textfont=scriptsize]{subfig}
\usepackage{textcomp}
\usepackage{stfloats}
\usepackage{url}
\usepackage{makecell}
\usepackage{verbatim}
\usepackage{graphicx}
\usepackage{booktabs}
\usepackage{longtable}
\usepackage{array}
\usepackage{cite}
\usepackage{xcolor}
\renewcommand{\algorithmicrequire}{\textbf{Input:}}

\begin{document}

\title{Multi-User MISO with Stacked Intelligent Metasurfaces: A DRL-Based Sum-Rate Optimization Approach}
\author{Hao Liu, \IEEEmembership{Graduate Student Member,~IEEE,} Jiancheng An, \IEEEmembership{Member,~IEEE,}\\ George C. Alexandropoulos, \IEEEmembership{Senior Member,~IEEE,} Derrick Wing Kwan Ng, \IEEEmembership{Fellow,~IEEE}, \\Chau Yuen, \IEEEmembership{Fellow,~IEEE}, \and and Lu Gan, \IEEEmembership{Member,~IEEE}

\thanks{This work is partially supported by National Natural Science Foundation of China under Grant 62471096. This research is supported in part by the Ministry of Education Tier 2 under Grant T2EP50124-0032, and Infocomm Media Development Authority under Grant FCP-NTU-RG-2024-025.

H. Liu and L. Gan are with the School of Information and Communication Engineering, University of Electronic Science and Technology of China (UESTC), Chengdu, Sichuan 611731, China. L. Gan is also with the School of Information and Communication Engineering, the Yibin Institute of UESTC, Yibin, Sichuan 644000, China (e-mail: liu.hao@std.uestc.edu.cn, ganlu@uestc.edu.cn).

J. An and C. Yuen are with the School of Electrical and Electronics Engineering, Nanyang Technological University, Singapore 639798 (e-mail: jiancheng.an@ntu.edu.sg, chau.yuen@ntu.edu.sg).

G. C. Alexandropoulos is with the Department of Informatics and Telecommunications,
National and Kapodistrian University of Athens, 15784 Athens, Greece (e-mail: alexandg@di.uoa.gr).

D. W. K. Ng is with the School of Electrical Engineering and Telecommunications,
University of New South Wales (UNSW), Sydney, NSW 2052, Australia (e-mail: w.k.ng@unsw.edu.au).

This paper has been presented in part at the IEEE International Conference on Communications (ICC), Denver, USA, 2024 \cite{liu2024drl}.}\vspace{-0.5cm}}

\maketitle

\begin{abstract}
Stacked intelligent metasurfaces (SIMs) represent a novel signal processing paradigm that enables over-the-air processing of electromagnetic waves at the speed of light. Their multi-layer architecture exhibits customizable computational capabilities compared to conventional single-layer reconfigurable intelligent surfaces and metasurface lenses. In this paper, we deploy SIM to improve the performance of multi-user multiple-input single-output (MISO) wireless systems through a low complexity manner with reduced numbers of transmit radio frequency chains. In particular, an optimization formulation for the joint design of the SIM phase shifts and the transmit power allocation is presented, which is efficiently tackled via a customized deep reinforcement learning (DRL) approach that systematically explores pre-designed states of the SIM-parametrized smart wireless environment. The presented performance evaluation results demonstrate the proposed method’s capability to effectively learn from the wireless environment, outperforming both conventional precoding schemes and optimization algorithms. Furthermore, the implementation of hyperparameter tuning and whitening process significantly enhances the robustness of the proposed DRL framework.
\end{abstract}


\section{Introduction}

\subsection{Background}
\IEEEPARstart{T}{he} evolution of wireless networks strives to achieve higher transmission rates and lower network latency. With the advent of the fifth-generation (5G) mobile networks relying on advanced physical-layer technologies \cite{liu2024drl}, such as massive multiple-input multiple-output (MIMO) \cite{demir2022channel} and millimeter wave communications \cite{mmwave}, there has been a ten-fold improvement in terms of data rates \cite{tang2020wireless}. However, the emergence of the Internet-of-Everything (IoE) has imposed more stringent demands on wireless capacity \cite{tang2020wireless}, and current technologies can hardly, or cannot, satisfy the extremely high throughput and ultra-reliable low-latency service for swarms of massive heterogeneous devices. 

In wireless networks, multi-user interference significantly hinders system performance, impacting both quality of service (QoS) and spectral efficiency \cite{pan2022RIS_overview}. Consequently, managing interference remains a key challenge for next-generation networks, driving extensive research on advanced techniques \cite{hochwald2005vector, li2020tutorial, tang2020wireless, sun2018learning}. In particular, several key interference cancellation techniques have been developed over the past decades, including digital precoding schemes \cite{hochwald2005vector}, hybrid precoding approaches \cite{pan2022RIS_overview, tang2020wireless, nguyen2023secondaryRIS, huang2020reconfigurable_DRL}, methods based on artificial intelligence (AI) \cite{ zhang2022multiagent_ddqn, stylianopoulos2022icc, sun2018learning}, and more recently, fully analog frameworks based on the stacked intelligent metasurface (SIM) technology \cite{an2023stacked_6g, liu2022programmable, an2023stacked}.

\subsection{Related Works}

\subsubsection{Traditional Precoding Schemes} In practical multi-user systems, digital precoding schemes emerge as an efficacious tool for mitigating inter-user interference by exploiting accurate channel state information (CSI) \cite{liu2022deep_twc}. The effectiveness of precoding technology is evidenced by its spatial selectivity to mitigate multipath interference and extend coverage range. 
Among them, linear precoding schemes have received considerable research attention, offering an efficient way to linearly combine users' data signals while avoiding the computational burden associated with nonlinear approaches.
Specifically, these schemes include zero-forcing (ZF), minimum mean square error (MMSE), and signal-to-leakage-and-noise-ratio (SLNR) \cite{sadek2007leakage}. 
Yet, the implementation of advanced digital precoding techniques presents significant challenges for large-scale network deployment, primarily because they require high-precision digital-to-analog converters (DACs) and energy-hungry radio frequency (RF) chains, resulting in increased processing delays and power consumption.

\subsubsection{Reconfigurable Intelligent Surfaces} The aforementioned digital precoding techniques \cite{sadek2007leakage} 
addressed multi-user interference at the transceiver side. Recently, there has been increasing research and development interest in reconfigurable intelligent surfaces (RISs) \cite{TCOM_2025_An_Flexible, jia2023codebook, liu2023kmeans, TWC_2025_An_Flexible, mohamed2020ris_tccn}, which are ultra-thin planar structures composed of numerous cost-effective and nearly passive reflective units. RIS technology is intended to contribute to the establishment of favorable channels for next-generation communication networks with low power consumption \cite{mohamed2020ris_tccn, an2022codebook, an2022lowcomplexity, you2021ris_twc, yu2024environment_tcom}. Specifically, RISs can programmatically alter the phase of electromagnetic (EM) waves impinging upon them, without requiring any active RF chains or any form of power amplification \cite{TCOM_2025_An_Flexible}. Given these potential benefits, numerous studies have focused on exploiting RISs to mitigate interference via passive beamforming \cite{pan2022RIS_overview, tang2020wireless, nguyen2023secondaryRIS, cao2022massive}. For instance, by adapting the RIS phase shifts based on CSI, beams can be directed towards user equipments (UEs) while minimizing co-channel interference \cite{tang2020wireless}. Additionally, \emph{Zhou et al.} \cite{zhou2020IRS_MISO} proposed a robust beamforming method based on imperfect channels, further enhancing the versatility of RIS. 
However, these traditional approaches face significant challenges in scalability and computational efficiency, often falling short in delivering optimal performance under real-world dynamics.

\begin{table*}[!t]
\centering
\caption{A survey of metasurface-based wireless communication systems}\label{tab:survey}
\begin{tabular}{llllllll}
\toprule 
Reference & Scenario & Metasurface function & Programmable & Metric & Power allocation & Imperfect CSI & Algorithm \\
\midrule
\cite{sci} & - & Image classification & $\times$ & - & $\times$ & $\times$ & - \\
\cite{liu2022programmable} & - & Image classification & $\checkmark$ & - & $\times$ & $\times$ & -\\
\cite{an2023stacked_6g} & MIMO & Precoding/combining & $\checkmark$ & Fitting error & $\checkmark$ & $\times$ & GD\\
\cite{an2023stacked} & Multi-user MISO & Wave-based beamforming & $\checkmark$ & Sum rate & $\times$ & $\times$ & AO\\
\cite{lin2024stacked} & MISO satellite & Wave-based beamforming & $\checkmark$ & Sum rate & $\checkmark$ & $\times$ & AO \\
\cite{hassan2024efficient} & SISO & Wave-based beamforming & $\checkmark$ & Received power & $\times$ & $\times$ & AO \\
\cite{papazafeiropoulos2024achievable} & MIMO & Precoding/combining & $\checkmark$ & Achievable rate & $\checkmark$ & $\times$ & GD \\
\cite{Papazafeiropoulos2024statisticalCSI} & Multi-user MISO & Wave-based beamforming & $\checkmark$ & Sum rate & $\checkmark$ & $\times$ & AO \\
\cite{papazafeiropoulos2024near_field} & Multi-user MIMO & Wave-based beamforming & $\checkmark$ & Sum rate & $\checkmark$ & $\times$ & AO and BCD
\\
This paper & Multi-user MISO & Wave-based beamforming & $\checkmark$ & Sum rate & $\checkmark$ & $\checkmark$& DRL\\
\bottomrule 
\multicolumn{8}{l}{AO: Alternating optimization, GD: Gradient descent, BCD: Block coordinate descent}
\end{tabular}\vspace{-0.5cm}
\end{table*}

\subsubsection{Artificial Intelligence} Recently, AI techniques have garnered substantial interest for effective interference mitigation in multi-user systems \cite{xu2022deep_wcl, huang2020reconfigurable_DRL, sun2018learning, zhang2022multiagent_ddqn, stylianopoulos2022icc, alex2022ml_rl, liu2024drl}. In \cite{sun2018learning}, \emph{Sun et al.} customized a deep neural network (DNN) to mimic a weighted MMSE algorithm, achieving low complexity and efficient interference cancellation. Also, \emph{Xiao et al.} \cite{xiao2020RL_dense_cell} proposed a deep reinforcement learning (DRL)-based power control scheme for suppressing downlink interference while conserving energy in ultra-dense small cells. 
This highlights the inherent adaptability and model-free learning capability of DRL, making it particularly suitable for complex and dynamic wireless communication systems.
Furthermore, a multi-agent dual-depth Q-network-based approach was proposed in \cite{zhang2022multiagent_ddqn} to jointly optimize the beamforming vectors and power splitting ratios in a multi-user multiple-input single-output (MISO) simultaneous wireless information and power transfer system. Meanwhile, AI can be combined with RISs to enhance interference cancellation capabilities. Specifically, \emph{Huang et al.} \cite{huang2020reconfigurable_DRL} developed a DRL-based scheme to jointly optimize RIS and transmit beamforming at the base station (BS) in a multi-user MISO system. Moreover, \emph{Stylianopoulos et al.} \cite{stylianopoulos2022icc} proposed a multi-armed bandits DRL approach for quantized RIS in a multi-user MISO system, while \emph{Alexandropoulos et al.} \cite{alex2022ml_rl} elaborated on the design and necessary measurement requirements for DRL RIS controllers. \emph{Wang et al.} \cite{wang2024madrl} developed a multi-agent DRL framework and achieved near-optimal performance with substantially reduced computational complexity compared to the traditional AO method, highlighting the evolving role of AI in refining communication efficiencies.

\subsubsection{Stacked Intelligent Metasurfaces}
While AI techniques offer superior resilience to interference in large-scale networks and complex environments compared to digital precoding techniques such as MMSE and ZF, it is worth noting that DNN-based interference cancellation technologies still rely on sophisticated digital signal processors. More recently, inspired by the wave-based computation potential \cite{ICASSP_2025_Lin_UAV, TWC_2025_An_Stacked, yang2022next, an2024emerging}, an EM neural network (EMNN) was introduced \cite{sci}. This physical EMNN consists of several passive diffractive layers, enabling it to compute various complex functions and process input waves at the speed of light. This innovation substantially reduces energy consumption and remarkably shortens the processing delay compared to conventional neural networks relying on digital computing \cite{hu2024cellfree, niu2024isac}. Nevertheless, the EMNN is made of three-dimensional (3D) printed neurons, which are inherently constrained to solving a single specific task once fabricated. To address this limitation, a programmable EMNN, also known as SIM, was reported in \cite{liu2022programmable} by utilizing several reconfigurable metasurfaces, which can dynamically adapt the network coefficients in real-time through field-programmable gate arrays (FPGA). Indeed, SIM not only inherits the advantages of EMNN for network operations at the speed of light but also offers the high flexibility to be adaptively re-trained for accommodating a wide spectrum of machine learning tasks \cite{an2024sim_wc, huang2024semantic}.

With its computational power in the wave domain and adaptability to the environment, SIM has been proven its great potential to enhance performance across various wireless sensing and communication systems \cite{liu2024stackedintelligentmetasurfaceswireless, an2023stacked_DOA, yao2024channel, niu2024sim_siso}.
Specifically, \emph{Lin et al.} \cite{lin2024stacked} explored the application of SIM in low earth orbit satellite communication systems. \emph{Li et al.} \cite{li2024stacked} integrated a SIM with each BS in a cell-free communication system and leveraged AO to optimize active beamforming and SIM phase shifts. Moreover, in a near-field communication system, \emph{Papazafeiropoulos et al.} \cite{papazafeiropoulos2024near_field} utilized AO and BCD techniques to jointly optimize the transmit power and SIM phase shifts. 
\emph{An et al.} \cite{an2023stacked} and \emph{Papazafeiropoulos et al.} \cite{Papazafeiropoulos2024statisticalCSI} explored the application of SIM to multi-user beamforming in an MISO system and presented an AO approach to address the non-convex optimization problem. In particular, the AO algorithm alternately tackles the power allocation and SIM phase shift design subproblems via iterative water-filling and GD approaches, respectively \cite{an2023stacked}.

\subsection{Contribution}

However, conventional AO schemes, which optimize variables alternately, tend to converge to locally optimal solutions since they decompose the joint optimization problem into several subproblems. Furthermore, it also results in prohibitive computational complexity for optimizing the numerous meta-atoms in the SIM, thereby complicating adaptation to dynamic wireless environments. Moreover, the GD method necessitates calculating the gradient formula based on the specific optimization problem, thus lacking generalizability. In contrast, DRL represents a promising paradigm independent of the problem's specific form. It enables wireless systems to acquire knowledge of complex, dynamic wireless environments without needing large-scale data like traditional neural network-based methods. This is accomplished through continuous self-learning from network feedback \cite{alex2022ml_rl}. Motivated by the aforementioned potential, DRL is increasingly recognized as an effective approach to acquire near-optimal solutions for complex joint optimization and decision-making problems, while maintaining relatively low computational complexity.

Building on these insights, in this paper, we present a SIM-assisted multi-user downlink MISO communication system optimized by DRL. To elaborate, we contrast this work with related SIM-aided wireless communications research at the time of writing in Table I. Specifically, our contributions are summarized as follows:
\begin{enumerate}
 \item We consider a SIM-assisted multi-user MISO wireless communication system. The SIM, which consists of a group of metasurfaces, enables efficient signal processing in the wave domain at light speed and is seamlessly integrated with the BS. By leveraging this advanced and efficient computing paradigm, we can completely eliminate the use of a digital precoder and significantly reduce the required number of RF chains at the BS.
 
 \item We formulate an optimization problem that jointly optimizes the transmission coefficients of the meta-atoms in the SIM and the power allocation strategy for the transmit antennas, with the objective of maximizing the sum rate of the multi-user MISO system. However, the problem is non-convex in nature due to the constraints imposed by the passive units of the SIM. 
 
 \item We propose a DRL-based method to address the challenge of jointly optimizing wave-based beamforming and power allocation. 
We meticulously design the deep deterministic policy gradient (DDPG) architecture, incorporating a carefully crafted attenuated whitening process to enhance its effectiveness in tackling intricate optimization problems within continuous solution spaces. Furthermore, this framework integrates a residual convolutional neural network (CNN), augmented with a multi-branch network specifically tailored to the unique SIM structure.
The DRL technique precludes explicit prior training data collection, thereby obviating the need for time-consuming labeled data collection for algorithm training. Furthermore, the computational complexity of the proposed DRL method is thoroughly analyzed.
 
 \item Numerical results demonstrate the advantages of the proposed SIM-assisted multi-user MISO wireless communication system across various practical scenarios. These findings corroborate the enhanced sum rate improvement of the joint optimization scheme based on DRL in the wireless communication network.
\end{enumerate}

The rest of the paper is organized as follows. The SIM-assisted multi-user MISO system model and optimization problem formulation are described in Section II. The proposed DRL-based framework and the parameter updating process of DDPG are presented in Section III. The simulation results and the corresponding analysis are presented in Section IV to verify the performance of the proposed algorithms. Finally, conclusions are provided in Section V.

\emph{Notations}: Bold lowercase and uppercase letters denote vectors and matrices, respectively; $\mathbb{E}(\cdot)$ denotes the expectation operation; $(\cdot)^T$ and $(\cdot)^H$ represent the transpose and Hermitian transpose, respectively; $\Re (x)$ and $\Im (x)$ represent the real and imaginary parts of complex-valued $x$, respectively; $\text{diag}(\mathbf{c})$ denotes a diagonal matrix with the elements of vector $\mathbf{c}$ on its main diagonal; $[\mathbf{X}]_{n,m}$ denotes the $n$-th row and $m$-th column element of $\mathbf{X}$; $[\mathbf{X}]_{n,:}$ and $[\mathbf{X}]_{:,m}$ denote vectors collecting the elements of the $n$-th row and $m$-th column of matrix $\mathbf{X}$, respectively; $\mathbb{C}^{x \times y}$ denotes the space of $x \times y$ complex-valued matrices; $\lceil x \rceil$ means the nearest integer greater than or equal to $x$; $\text{mod}(x,y)$ represents the remainder after division of $x$ by $y$; $\mathbf{g} \sim \mathcal{CN}(\mathbf{0}, \mathbf{\Sigma})$ denotes the distribution of a circularly symmetric complex Gaussian (CSCG) random vector $\mathbf{g}$ with zero mean and covariance $\mathbf{\Sigma}$, where $\sim$ means ``distributed as"; $\mathbf{I}$ represents a unit matrix; $\text{log}_a$ denotes the logarithmic function with base $a$; $|x|$ and $\lVert \mathbf{x} \rVert $ represent the magnitude of a complex number $x$ and the Euclidean norm of vector $\mathbf{x}$, respectively; $\otimes$ represents the Kronecker product; $\text{sinc}(x) = \text{sin} (\pi x) / (\pi x)$ represents the sinc function. $\mathcal{O}(\cdot)$ represents the computational complexity order.

\section{Modeling and Problem Formulation}
This section introduces the SIM-assisted multi-user MISO system model, wherein the SIM performs wave-based precoding. Specifically, Section II-A first presents the proposed SIM design. Then, Section II-B describes the spatially correlated channel model for the multi-user MISO system. Section II-C subsequently outlines the system performance metric and Section II-D formulates the design as an optimization problem.
\subsection{SIM Design}
\begin{figure*}[!t]
\centering
\includegraphics[width=0.8\linewidth]{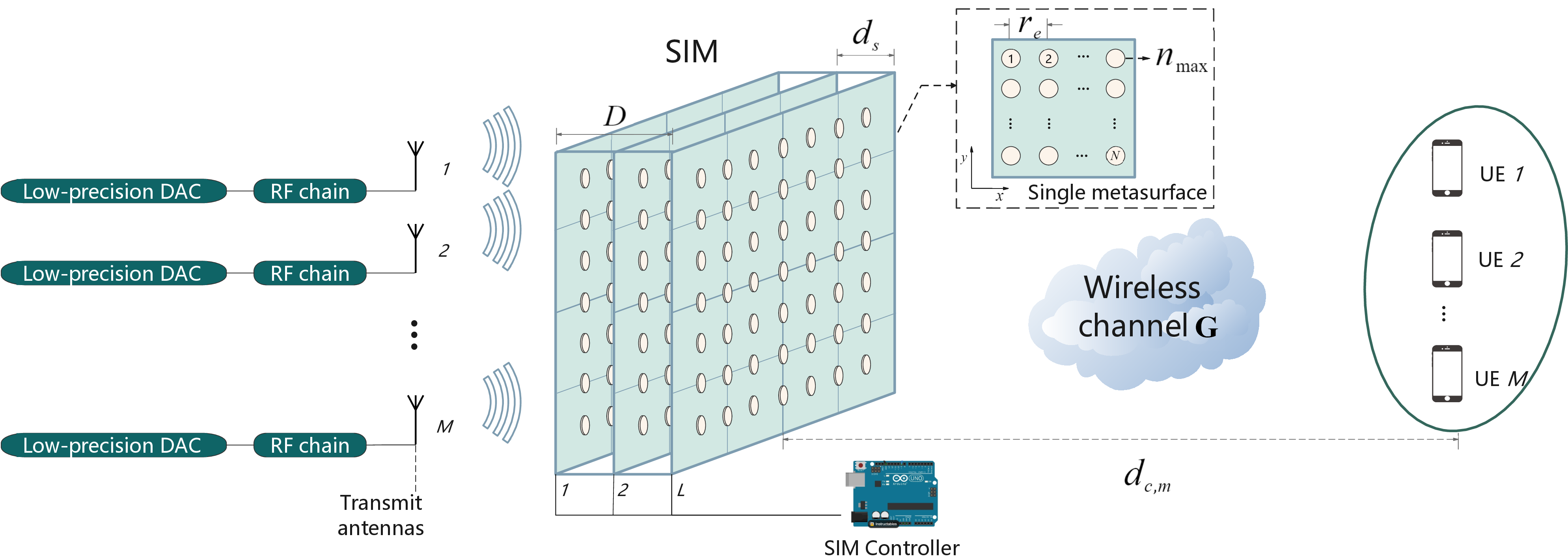}
\caption{The considered SIM-aided multi-user MISO transmission system with $L$ layers of metasurfaces.}\vspace{-0.5cm}
\label{fig1}
\end{figure*}

As shown in Fig. \ref{fig1}, we consider a downlink MISO system comprised of a BS, a SIM, and multiple UEs, in which the SIM is integrated with the BS to improve system performance. The SIM consists of multiple programmable metasurfaces, each of which is composed of a large number of meta-atoms that are linked to an intelligent controller, e.g., an FPGA \cite{liu2022programmable}. By appropriately tuning the transmission coefficients of the SIM via this controller, the SIM is capable of directly manipulating the electromagnetic (EM) behaviors of the propagating waves \cite{sci}. As a result, SIM can tailor a spatial waveform shape at the output metasurface layer. Note that traditional communication systems without SIM require many active RF chains to generate the desired EM waveforms. In contrast, employing SIM can significantly reduce the number of active RF chains \cite{an2023stacked} and enable low-precision DACs\footnote{In practical communication systems, low-precision DACs can be utilized by employing modulation schemes with lower DAC requirements, such as binary phase-shift keying (BPSK). This approach accordingly reduces overall system power consumption and cost.}. This is achieved by directly steering the EM waves as they pass through the low-cost and energy-efficient metasurfaces.

Let $M$ represent the number of UEs with the corresponding set given by $\mathcal{M} = \{1,2,..., M\}$. At the BS, $M$ antennas are selected to concurrently transmit $M$ individual data streams associated with $M$ UEs\footnote{In this paper, we assume that $M$ antennas have been selected. The task of simultaneously optimizing antenna selection and wave-based precoding constitutes future research work.}.
Let $L$ denote the number of equidistance-spaced metasurface layers of the SIM and $N$ denote the number of meta-atoms on each metasurface layer, satisfying $N \geq M$, while the metasurface layer set is represented by $\mathcal{L} = \{1,2,..., L\}$ and the meta-atom set of each layer is represented by $\mathcal{N}=\{1,2,..., N\}$. Besides, let $\phi_n^l=e^{j\varphi_n^l}$ denote the transmission coefficient of the $n$-th meta-atom on the $l$-th metasurface layer. $\varphi_n^l$ denotes the corresponding phase shift, which is assumed to be adjustable between $0$ and $2\pi$ continuously \cite{an2023stacked, sci, an2023stacked_6g}, i.e., $\varphi_n^l \in [0, 2\pi)$, $n\in\mathcal{N}$, $l\in \mathcal{L}$. Then, the transmission coefficient vector and the corresponding matrix representation of the $l$-th metasurface are denoted by $\boldsymbol{\phi}^l=[\phi_1^l, \phi_2^l, ..., \phi_N^l]^T \in \mathbb{C}^{N\times1}$ and $\boldsymbol{\Phi}^l = \text{diag}(\boldsymbol{\phi}^l)\in \mathbb{C}^{N\times N}$, respectively.

Without loss of generality, each metasurface layer of the SIM is modeled as a uniform planar array, with all the metasurface layers being arranged in an identical square configuration. Let $r_e$ denote the element spacing between two adjacent meta-atoms within the same layer. Thus, the spacing between the $n$-th meta-atom and the $\tilde{n}$-th meta-atom on the same layer is
\begin{equation}
\label{space_same_layer}
r_{n, \tilde{n}} = r_e \sqrt{(n_x - \tilde{n}_x)^2 + (n_y - \tilde{n}_y)^2},
\end{equation}
where $n_x$ and $n_y$ are the $x$-axis and $y$-axis indices of the $n$-th meta-atom, respectively, with $\tilde{n}_x$ and $\tilde{n}_y$ defined similarly for the $\tilde{n}$-th meta-atom. The indices are calculated as $n_x = \text{mod} (n-1,\, n_{\mathrm{max}}) + 1$ and $n_y = \lceil n / n_{\mathrm{max}} \rceil$, where $n_{\mathrm{max}}=\sqrt{N}$ represents the maximum number of meta-atoms per row.

Furthermore, the total thickness of the SIM structure is denoted as $D$, where the parallel metasurface layers are equally spaced with a distance of $d_s = D/(L-1)$ between adjacent layers. Thus, the spacing between the $n$-th meta-atom on the $l$-th layer and the $\tilde{n}$-th meta-atom on the ($l-1$)-th layer is $r_{n, \tilde{n}}^l = \sqrt{r_{n, \tilde{n}}^2 + d_s^2 }$.
Moreover, we assume that the transmit antennas are arranged in a uniform linear array (ULA) with a half-wavelength distance between them. The array's center aligns with that of the first metasurface. Thus, the distance between the $m$-th transmitting antenna and the $n$-th meta-atom on the first metasurface is defined as
\begin{align}
 \label{space_first_layer}
 r_{m,n}^1 &= \bigg ( \Big [\Big (n_y - \frac{n_{\mathrm{max}} + 1}{2} \Big ) r_e - \Big (m -
 \frac{M+1}{2} \Big ) 
 \frac{\lambda}{2} \Big ]^2 \nonumber \\
 & \quad + \Big (n_x - \frac{n_{\mathrm{max}}+1}{2} \Big )^2 r_e^2 + d_s^2 \bigg )^{\frac{1}{2}},
\end{align}
where $\lambda$ denotes the wavelength of the carrier signal. 

Furthermore, the propagation coefficient between adjacent metasurface layers can be obtained from the Rayleigh-Sommerfeld diffraction equation\footnote{Note that the Rayleigh-Sommerfeld diffraction equation inherently accounts for the inter-layer power loss of the EM wave propagation between metasurface layers.} \cite{sci, an2023stacked}. According to these equations, the propagation coefficient from the $\tilde{n}$-th meta-atom on the $(l-1)$-th metasurface layer to the $n$-th meta-atom on the $l$-th metasurface layer is defined by
\begin{equation}
 \label{coff}
 [\mathbf{W}^l]_{n, \tilde{n}} = \frac{d_s s_a}{(r_{n, \tilde{n}}^l)^2} \left( \frac{1}{2 \pi r_{n, \tilde{n}}^l} - j\frac{1}{\lambda} \right) e^{j2\pi r_{n,\tilde{n}}^l / \lambda}, 
\end{equation}
$\mathrm{for}\, l\in\mathcal{L}/\{1\},\,n,\tilde n \in \mathcal{N}$, where $s_a$ represents the area of each meta-atom. The propagation coefficient $[\mathbf{W}^1]_{n,m}$ from the $m$-th transmitting antenna to the $n$-th meta-atom on the first metasurface is obtained by replacing $r_{n, \tilde{n}}^l$ in (\ref{coff}) with $r^1_{m,n}$ in (\ref{space_first_layer}). Furthermore, we assume that non-adjacent layers do not interact with each other \cite{an2024sim_wc, liu2024stackedintelligentmetasurfaceswireless, an2023stacked_DOA, yao2024channel, li2024stacked, an2023tutoral, nerini2024physically}. Thus, the overall EM response in the SIM can be expressed as
\begin{equation}
 \label{SIM_matrix}
 \mathbf{B} = \mathbf{\Phi}^L \mathbf{W}^L \cdots \mathbf{\Phi}^2 \mathbf{W}^2 \mathbf{\Phi}^1 \mathbf{W}^1 \in \mathbb{C}^{N\times M}.
\end{equation}

\subsection{Spatially Correlated Channel Model}
For the wireless channels shown in Fig. \ref{fig1}, we consider the Rician channel model.
Specifically, the channel spanning from the output metasurface to the $m$-th UE can be expressed as
\begin{equation}
\mathbf{g}_m^H = \sqrt{\rho_m / (R_H + 1)}(\sqrt{R_H} \mathbf{g}^H_{m,\mathrm{LoS}} + \mathbf{g}^H_{m,\mathrm{NLoS}}) \in \mathbb{C}^{1 \times N},
\end{equation}
where $\mathbf{g}^H_{m,\mathrm{LoS}}$ and $\mathbf{g}^H_{m,\mathrm{NLoS}}$ represent the line-of-sight (LoS) and non-LoS (NLoS) components of the $m$-th UE channel $\mathbf{g}_m^H$, respectively. $R_H$ represents the Rician factor. $\rho_m$ denotes the path loss of the $m$-th UE channel, given by 
\begin{equation}
 \label{path_loss}
 \rho_m = C_0 d_{c,m}^{-\alpha},\; m \in \mathcal{M},
\end{equation}
where $C_0 \in \mathbb{R}$ denotes the path loss at a reference distance of $1$ meter, $d_{c,m}$ denotes the transmission distance between the SIM and the $m$-th UE, and $\alpha$ denotes the corresponding path loss exponent.
Due to the closely spaced meta-atoms, we consider a spatially correlated channel model \cite{an2024sim_wc, liu2024stackedintelligentmetasurfaceswireless, an2023stacked_DOA, yao2024channel, li2024stacked, an2023tutoral, nerini2024physically} to characterize the NLoS component of $\mathbf{g}_m^H$, yielding 
\begin{equation}
 \mathbf{g}^H_{m,\mathrm{NLoS}} = [\Tilde{\mathbf{g}}^H_{m,\mathrm{NLoS}}]_{m,:} \mathbf{R}^{1/2} \in \mathbb{C}^{1 \times N},
\end{equation}
where $[\Tilde{\mathbf{g}}^H_{m,\mathrm{NLoS}}]_{m,:} \sim \mathcal{CN}(\mathbf{0}, \mathbf{I}_{1\times N})$ and $\mathbf{R} \in \mathbb{C}^{N \times N}$ denotes the spatial correlation matrix of the SIM. Assuming far-field wave propagation in an isotropic scattering wireless environment, the spatial correlation matrix $\mathbf{R}$ is thus defined by \cite{an2023tutoral, demir2022channel}
\begin{equation}
 \label{correlation}
 [\mathbf{R}]_{n, \tilde{n}} = \text{sinc}(2r_{n, \tilde{n}} / \lambda),\; n \in \mathcal{N},\; \tilde{n} \in \mathcal{N}.
\end{equation}
Moreover, let $\psi_m \in [0, 2\pi)$ and $\beta_m \in [0, \pi/2]$ represent the physical azimuth and elevation angle of departure (AoD) or angle of arrival (AoA) of the $m$-th UE, respectively. Consequently, the electrical angles $\vartheta_{x,m}$ and $\vartheta_{y,m}$ along the $x$ and $y$ axes are given by $\vartheta_{x,m} = \frac{2\pi}{\lambda} r_e \sin(\beta_m) \cos(\psi_m)$ and $\vartheta_{y, m} = \frac{2\pi}{\lambda} r_e \sin(\beta_m) \sin(\psi_m)$, respectively. Thus, the steering vector of the last metasurface layer $\mathbf{a}_m(\vartheta_{x,m}, \vartheta_{y,m}) \in \mathbb{C}^{1 \times N}$ is expressed as 
\begin{equation}
\begin{split}
 \mathbf{a}_m(\vartheta_{x,m}, \vartheta_{y,m}) = [1, e^{j\vartheta_{x,m}}, ..., e^{j\vartheta_{x,m} n_{\mathrm{max}}}] \\ 
 \otimes [1, e^{j\vartheta_{y,m}}, ..., e^{j\vartheta_{y,m} n_{\mathrm{max}}}].
\end{split}
\end{equation}
Hence, the LoS component of $\mathbf{g}_m$ is defined as follows:
\begin{equation}
 \mathbf{g}^H_{m,\mathrm{LoS}} = \mathbf{a}_m(\vartheta_{x, m}, \vartheta_{y, m}) \in \mathbb{C}^{1 \times N}.
\end{equation}

\subsection{System Performance Metric}
We focus on the downlink data transmission and the signal received at the $m$-th UE is given as
\begin{equation}
 \label{receive_signal}
 y_m = \mathbf{g}_m^H \mathbf{B} \mathbf{P} \mathbf{x} + z_m,\,\forall m,
\end{equation}
where $z_m$ denotes the additive white Gaussian noise (AWGN) with variance $\sigma^2_m$, i.e., $z_m \sim \mathcal{CN}(0, \sigma^2_m)$. $\mathbf{P}=\text{diag}(\sqrt p_1, \sqrt p_2, ..., \sqrt p_M) \in \mathbb{C}^{M\times M}$ denotes the transmission power matrix with $p_m, m\in \mathcal{M}$ representing the transmit power allocated to the $m$-th UE. $\mathbf{x}=[x_1, x_2, ..., x_M] \in \mathbb{C}^{M \times 1}$ denotes a column vector of the data streams transmitted to all the UEs, satisfying $\mathbb{E}[\lVert \mathbf{x} \rVert^2] = 1$. We also assume that our passive SIM does not introduce any noise to its internally propagating signals \cite{liu2022programmable}.

Furthermore, the received signal of (\ref{receive_signal}) can be rewritten as
\begin{equation}
 \label{receive_signal_rewrite}
 y_m = \sqrt{p_m} \mathbf{g}_m^H [\mathbf{B}]_{:,m} x_m 
 + \sum^M_{k=1,k \neq m} \sqrt{p_k} \mathbf{g}_m^H [\mathbf{B}]_{:,k} x_k 
 + z_m.
\end{equation}
The second addend of (\ref{receive_signal_rewrite}) is considered as the co-channel interference caused by other $(M-1)$ users. Thus, the received signal-to-interference-plus-noise ratio (SINR) at the $m$-th UE is defined as 
\begin{equation}
 \label{sinr}
 \kappa_m = \frac{p_m |\mathbf{g}_m^H [\mathbf{B}]_{:,m}|^2}{
 \sum^M_{k =1,k \neq m} p_k |\mathbf{g}_m^H [\mathbf{B}]_{:,k}|^2 + \sigma_m^2}.
\end{equation}
In this paper, we adopt the sum rate of $M$ UEs as a system performance metric, which is defined by
\begin{equation}
 \label{equ:sum_rate}
 C(\mathbf{P}, \boldsymbol{\Phi}^l) = \sum^M_{m=1} \text{log}_2 (1+\kappa_m).
\end{equation}

\subsection{Problem Formulation}
 Our objective is to design the SIM phase shifts $\mathbf{\Phi}^l$ and the transmit power allocation strategy $\mathbf{P}$ that jointly maximize the system sum-rate $C(\mathbf{P}, \boldsymbol{\Phi}^l)$ under perfect CSI of all involved wireless channels. As such, the joint design can be formulated as the following joint optimization problem:
 
\begin{subequations}
\label{problem}
\begin{align}
\!\!(\text{P1}): \max_{\{\boldsymbol{\Phi}^l\}_{l=1}^L, \{p_m\}_{m=1}^M } & C(\mathbf{P}, \boldsymbol{\Phi}^l) \label{problem_1} \\
 \mathrm{s.t.} \quad \quad & \mathbf{B} = \mathbf{\Phi}^L \mathbf{W}^L \cdots \mathbf{\Phi}^2 \mathbf{W}^2 \mathbf{\Phi}^1 \mathbf{W}^1 \label{problem_3}, \\
 & |\boldsymbol{\phi}^l_n| =1,\; l\in \mathcal{L}, \; n \in \mathcal{N} \label{problem_2}, \\
 &\mathbf{P} = \text{diag}(\sqrt{p_1}, \sqrt{p_2}, ..., \sqrt{p_M}), \label{constraintd}\\
 & \sum_{m=1}^{M} p_m \leq P, \label{constrainte}\\
 & p_m \geq 0, \label{problem_4}
\end{align}
\end{subequations}
where $P$ denotes the maximum transmit power budget at the BS. 
It can be verified that problem (P1) constitutes non-convex optimization due to non-convexities arising from the objective function, constraints (\ref{problem_2}), and consequently, in the multilayered architecture of the SIM (\ref{problem_3}). 
Traditional optimization techniques, such as AO methods \cite{wu2019RIS_beamforming, guo2020weighted_sum_rate, an2023stacked}, often struggle to find satisfactory solutions within moderate complexity due to the large number of tunable elements introduced by its multilayer structure and the continuous phase space. The GD methods require the reduplicative computation of gradient for specific optimization problems, lacking generalizability in wireless systems. Exhaustive search approaches, while theoretically capable of finding an optimal solution, become computationally intractable given the high-dimensional parameter space. Moreover, conventional neural network-based methods require extensive pre-collected training data, limiting adaptability to varying system conditions.

To overcome these challenges, we propose a novel DRL-based approach that directly optimizes the phase shifts of the SIM, $\mathbf{\Phi}^l,\,l\in \mathcal{L}$ and transmit power allocation $\mathbf{P}$ with low computational complexity. Unlike traditional deep learning algorithms that rely on both offline training and online inference \cite{franccois2018introduction, huang2020reconfigurable_DRL}, our DRL framework continuously interacts with the wireless environment, enabling self-guided exploration and learning from real-time feedback and eliminating the need for exhaustive data collection, which makes it particularly well-suited for large-scale, dynamically evolving wireless communication systems.

\section{Proposed DRL Design}
In this section, we first introduce the fundamental framework of DRL, detailing the composition of the state, action, reward, and policy, to establish the foundation for the proposed optimization method. Subsequently, the entire network architecture is presented, followed by a detailed explanation of the specific optimization process. Finally, we analyze the computational complexity of the proposed DRL-based scheme.

\subsection{DRL Formulation}
Model-free DRL is a dynamic tool capable of solving a series of decision-making problems, enabling an agent to learn the best strategy in a real-time manner \cite{franccois2018introduction, xiao2020RL_dense_cell, zhang2022multiagent_ddqn}. A general DRL system consists of two components: the agent and the environment. The agent continuously interacts with the environment, applying actions based on its current policy and observing the immediate rewards and new states from the environment. The fundamental concepts of the proposed DRL formulation for handling (P1) are as follows:
\begin{itemize}
 \item \emph{Action space}: Let $\mathcal{A}$ denote the action space. According to the state of the environment $\mathbf{s}_t$, the agent takes an action $\mathbf{a}_t \in \mathcal{A}$ to influence the environment according to a policy. Then, the agent receives the reward $r_t$ and observes the new state $\mathbf{s}_{t+1}$ feedback. For our system model, the action at each time step $t$ is constructed by the current SIM phase shifts $\boldsymbol{\phi}_t^l, \,l\in \mathcal{L}$ and transmit power allocation $p_{m,t},\,m\in \mathcal{M}$. Since DNNs require real-valued inputs and outputs, the SIM phase shifts are rearranged into real and imaginary components, i.e., 
 \begin{equation}
 \mathbf{a}_t = \big [ \Re( \boldsymbol{\phi}^1_t ), ..., \Im( \boldsymbol{\phi}^1_t ), ..., \Im( \boldsymbol{\phi}^L_t ), 
 p_{1,t}, ..., p_{M,t} \big ]^T,
 \end{equation}
 where $p_{m,t}$ is the transmit power of the $m$-th antenna at time step $t$. Hence, the action is represented as a vector with dimension $D_a = 2NL + M$.

 \item \emph{State space}: 
 Let $\mathcal{S}$ denote the environment state space, which encapsulates the essential information required for the DRL agent's decision-making process from the environment. The current state $\mathbf{s}_t \in \mathcal{S}$ observed at each time step $t$ consists of three components: the reward $r_{t-1}$ and action $\mathbf{a}_{t-1}$ from the previous time step, as well as the current CSI for all UEs, i.e.,
 \begin{equation}
 \begin{aligned} 
 \mathbf{s}_t = \big [ r_{t-1}, \mathbf{a}_{t-1}, \Re(\mathbf{g}^H_1),..., \Im(\mathbf{g}^H_1),...,\Im(\mathbf{g}_M^H) \big ]^T.
 \end{aligned}
 \end{equation}
Hence, the dimension of the state space is $D_s = 2N(L+M)+M+1$. 
The inclusion of $r_{t-1}$ and $\mathbf{a}_{t-1}$ provides the agent with historical context, enabling it to better capture temporal dependencies in the optimization process. Since DRL relies on iterative updates to approximate the optimal policy, incorporating past actions and rewards also helps mitigate policy fluctuations and improves learning stability, especially in high-dimensional continuous action spaces. Meanwhile, the current CSI ensures that the agent remains adaptive to real-time network dynamics.

 \item \emph{Policy}: Let $\pi$ denote the policy of agent and $\pi(\mathbf{s}_t, \mathbf{a}_t)$ denote the probability of taking $\mathbf{a}_t$ based on the environment state $\mathbf{s}_t$, satisfying $\sum_{\mathbf{a}_t \in \mathcal{A}} \pi(\mathbf{s}_t,\mathbf{a}_t) = 1$.
 
 \item \emph{Reward}: 
 The reward $r_t$ serves as a measure to quantify the quality of the policy and guides the agent in optimizing its decisions. Specifically, the agent takes action $\mathbf{a}_t$ in the environment and receives the reward as feedback, which reflects the system's performance under the current power allocation scheme $\mathbf{P}$ and SIM phase shift matrix ${\boldsymbol{\Phi}^l}$. The agent then dynamically adjusts its policy to maximize this reward over successive iterations. The primary objective of this paper is to maximize the sum rate of $M$ UEs. As such, the reward at each time step $t$ is defined as $C(\mathbf{P}, \boldsymbol{\Phi}^l)$, representing the sum rate determined by the outputs of the actor training neural network, which jointly optimizes $\mathbf{P}$ and ${\boldsymbol{\Phi}^l}$ to maximize the system performance. This mechanism ensures that the agent aligns its actions with the system's performance goals, leveraging the reward to iteratively refine its policy for optimal results.

 \item \emph{Transitions}: Action $\mathbf{a}_t$ is sampled from $\pi(\mathbf{s}_t,\mathbf{a}_t)$ which determines the phase shifts of the SIM and the power allocation. A transmission then occurs within a coherence block where $\mathbf{g}_m^H$ remains constant and $C(\mathbf{P}, \boldsymbol{\Phi}^l)$ is computed. Then, the environment/system proceeds to time step $t+1$ where new channel matrices are sampled/observed in independent and identically distributed (i.i.d.) fashion.
\end{itemize}

Generally, there are two categories of algorithms to determine the policy in DRL, i.e., value-based algorithms and policy-based algorithms \cite{stylianopoulos2022icc, alex2022ml_rl}. Specifically, the well-known deep Q-network (DQN) is a quintessential value-based algorithm that employs experience replay and target network strategies to enhance the stability and efficiency of its learning process. However, DQN is typically applicable to a discrete action space. Yet, in the SIM-aided scenario, a continuous action space spanned by SIM phase shifts and the transmit power allocation strategy is presented in (\ref{problem}). 
On the other hand, the policy gradient (PG) method is a basic policy-based algorithm that aims to maximize the long-term expected discounted reward of each episode for continuous action spaces \cite{alex2022ml_rl}. In fact, the PG directly adjusts the likelihood of selecting particular actions based on the gradient of expected reward. However, the pure PG approach exhibits high variance, leading to unstable learning. Moreover, it demands an extensive sample set.
Thus, the convergence performance and stability of PG in the wireless communication environment need to be improved \cite{franccois2018introduction}. 
To combine the advantages of both value-based and policy-based methods while mitigating their respective limitations, the hybrid actor-critic (AC) algorithm emerges as a promising solution \cite{alex2022ml_rl}. In the AC framework, the actor updates the policy based on the critic's feedback to handle continuous action spaces. However, the basic AC algorithm may still suffer from training instability and sample inefficiency when dealing with complex continuous control tasks such as the SIM configuration in the considered.

\begin{figure}
 \centering
 \includegraphics[width=0.9\linewidth]{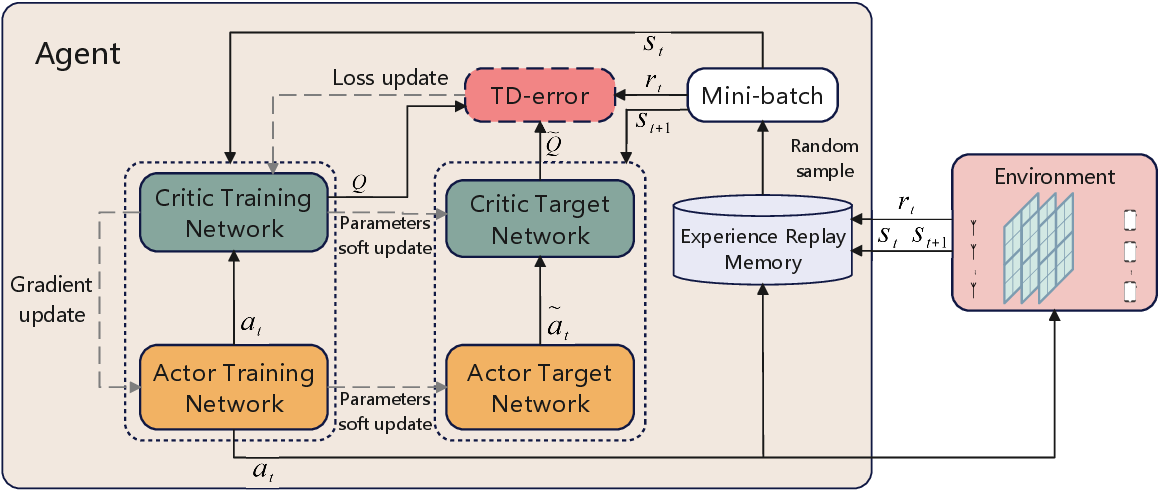}
 \caption{The proposed DRL-based SIM design framework adopting DDPG.}\vspace{-0.5cm}
 \label{fig:DDPG}
\end{figure}

Building upon the AC architecture, we adopt a DDPG-based solution, which enhances the basic AC framework with deep neural networks, experience replay, and target networks to achieve better sample efficiency and training stability. Thus, DDPG can handle continuous action space along with excellent stability and convergence, to solve the problem in (P1). DDPG also exploits a Q-function to choose an action $\mathbf{a}_t$ under the policy $\pi$ according to the current state $\mathbf{s}_t$. Specifically, the goal of DDPG is to find the best policy $\pi_{\mathrm{op}}$ that maximizes the long-term expected discounted reward. The Q-function evaluates state-action pairs via a Q-value, and the DDPG employs a separate critic as the Q-function. As shown in Fig. \ref{fig:DDPG}, the DDPG algorithm utilizes two neural networks with distinct roles: actor networks (depicted in orange) and critic networks (in green). The DDPG exploits the actor neural network $\pi(\mathbf{s}_t;\theta_\pi)$, with $\theta_\pi$ representing its parameters, as the policy to generate the SIM phase shifts and transmit power allocation strategy from a continuous action space $\mathcal{A}$. Besides, the critic neural network $Q(\mathbf{s}_t,\mathbf{a}_t;\theta_q)$, having the parameters $\theta_q$, is utilized to output a Q-value, which measures the output of the actor network. 

During the training phase, DRL operates without target data, which actually differentiates it from classic supervised training \cite{alex2022ml_rl, huang2020reconfigurable_DRL, sun2018learning}. To address this, the proposed DDPG adopts two networks with identical structures but with different parameters as shown in Fig. \ref{fig:DDPG}. The critic target network $\tilde Q(\mathbf{s}_{t+1},\tilde{\mathbf{a}}_t; \theta_{\tilde q})$ is utilized to update the critic training network $Q(\mathbf{s}_t, \mathbf{a}_t;\theta_q)$, while the actor target network $\tilde \pi(\mathbf{s}_{t+1}; \theta_{\tilde \pi})$ updates the actor training network $\pi(\mathbf{s}_t;\theta_\pi)$, where $\tilde{\mathbf{a}}_t$ denotes the output of actor target network $\tilde \pi(\mathbf{s}_{t+1}; \theta_{\tilde \pi})$. The training neural networks and the target neural networks are built with an identical structure but different parameters. While the training networks undergo training as classic DNNs, the target networks are untrainable and aim to provide a label for the training network. The parameters of the target neural network are updated periodically. At the end of each step, the updates on the critic and actor target neural networks are expressed as
\begin{eqnarray}
 \label{equ:target_update_c}
 \theta_{\tilde q} \gets (1-\eta_c) \theta_{\tilde q} + \eta_c \theta_{q},\\
 \label{equ:target_update_a}
 \theta_{\tilde \pi} \gets (1-\eta_a) \theta_{\tilde \pi} + \eta_a \theta_{\pi},
\end{eqnarray}
where $\eta_c \in (0,1)$ and $\eta_a \in (0,1)$ denote the learning rate for updating the critic and actor target neural network, respectively. The updates on the target neural networks can stabilize and accelerate the convergence of the DRL training process \cite{alex2022ml_rl}.

Moreover, the critic training neural network's update employs the gradient of a loss function. The label in the loss function consists of the reward $r_t$, which is returned by the SIM-aided multi-user MISO wireless environment, and the output of the critic target neural network $\tilde Q(\mathbf{s}_{t+1},\tilde{\mathbf{a}}_t; \theta_{\tilde q})$, which can be expressed as
\begin{equation}
 \label{equ:target_value}
 V(\mathbf{s}_t, \mathbf{a}_t) = r_t + \mu \tilde Q(\mathbf{s}_{t+1}, \tilde{\mathbf{a}}_t; \theta_{\tilde q}),
\end{equation}
where $\mu$ denotes a discount factor. In this paper, we employ the mean square error (MSE) loss function, which is defined as
\begin{equation}
 \label{equ:critic_loss}
 l(\theta_q) = \left ( V(\mathbf{s}_t,\mathbf{a}_t) - Q(\mathbf{s}_t, \mathbf{a}_t;\theta_q) \right )^2,
\end{equation}
where the error between $V(\mathbf{s}_t,\mathbf{a}_t)$ and estimated value $Q(\mathbf{s}_t,\mathbf{a}_t;\theta_q)$ is termed as the temporal difference (TD) error \cite{alex2022ml_rl}. Thus, the parameter update of the critic training neural network can be expressed as
\begin{equation}
 \label{equ:critic_update}
 \theta_q^{(t+1)} = \theta_q^{(t)} - \gamma_c \Delta_{\theta_q}l(\theta_q),
\end{equation}
where $\gamma_c > 0$ denotes the corresponding learning rate.

The update on the actor training neural network is defined through the policy gradient theorem \cite{alex2022ml_rl}, as follows
\begin{equation}
 \label{equ:actor_update}
 \theta_{\pi}^{(t+1)} = \theta_{\pi}^{(t)} - \gamma_a \Delta_\pi Q(\mathbf{s}_t, \pi(\mathbf{s}_t;\theta_\pi); \theta_q) \Delta_{\theta_\pi}\pi(\mathbf{s}_t;\theta_\pi),
\end{equation}
where $\gamma_a > 0$ denotes its learning rate. $\Delta_\pi Q(\mathbf{s}_t, \pi(\mathbf{s}_t;\theta_\pi); \theta_q)$ is the gradient of the critic training network regarding the output of the actor training neural network. $\Delta_{\theta_\pi}\pi(\mathbf{s}_t;\theta_\pi)$ represents the gradient of the actor training neural network with respect to its parameters $\theta_\pi$. Notably, the update process of the actor training neural network depends heavily on the gradient of the critic training neural network according to the current policy. This interaction ensures that the action policy is updated towards a direction that maximizes the long-term expected sum rate, as estimated through the Q-values.
\begin{figure*}[!th]
 \centering
 \includegraphics[width=0.95\linewidth]{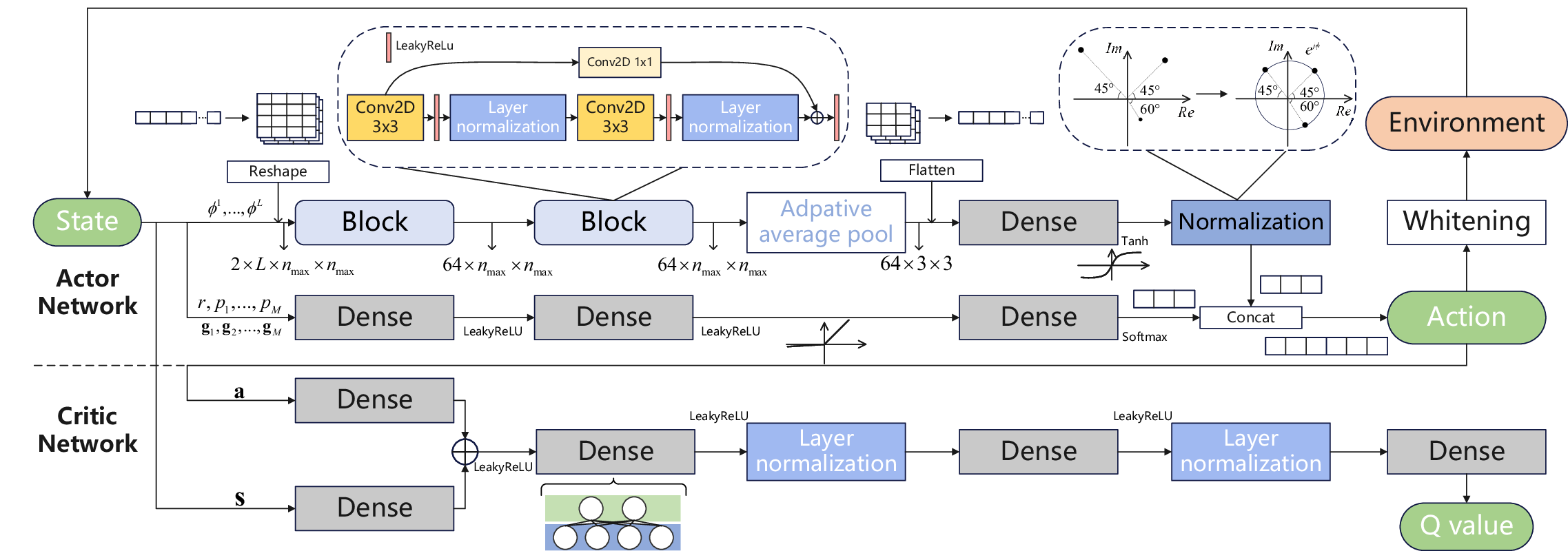}
 \caption{Architecture of the actor network and the critic network.}\vspace{-0.5cm}
 \label{fig:network}
\end{figure*}
It is noted, however, that value-based algorithms can be trapped in local optima due to the correlation between samples and nonstationary targets \cite{alex2022ml_rl, franccois2018introduction}. In addition, when considering short timescales, this behavior is particularly significant, since data generated through interactions with wireless environments tend to exhibit high temporal correlation. Therefore, the experience replay technique is utilized to reduce the negative impact of sample correlation on the agent by using a buffer window to store a portion of data. For each training phase, networks adopt a mini-batch randomly sampled from experience replay to calculate the gradient and update parameters. Moreover, in our DRL framework, we introduce noise to the action to prevent the learning process from being trapped in a locally optimal solution \cite{alex2022ml_rl}. This whitening process is applied to the SIM phase shifts and transmit power allocation prior to utilization within the wireless network. Specifically, truncated random white noise $\mathcal{W} \in [-w_a, w_a]$, where $w_a$ defines the truncation value, is integrated into the action processing, which is expressed as
\begin{equation}
 \label{equ:add_noise}
 \mathbf{a}_t \gets \mathbf{a}_t + \mathcal{W}.
\end{equation}
The noise $\mathcal{W}$ follows a Gaussian distribution with zero mean and variance $v$, i.e., $\widetilde{\mathcal{W}} \sim \mathcal{CN}(0, v)$. To enable smooth convergence, $v$ decreases exponentially during training as
\begin{equation}
 \label{equ:variance}
 v = v_0 \zeta^{t/t_{\mathrm{gap}}},
\end{equation}
where $v_0$ denotes the initial $v$ value while $\zeta \in (0,1)$ and $t_{\mathrm{gap}}$ are the decay rate and the gap factor for discounting the whitening process, respectively.

\subsection{Architecture of Employed Networks}
Fig. \ref{fig:network} shows the structures of the proposed networks. The actor neural network operates by inputting a state vector, directly producing the SIM phase shifts and transmit power allocation strategy. This network consists of two branches, each addressing the optimization of SIM phase shifts and power allocation, respectively. The branch assigned to process SIM phase shifts consists of a reshape operation, two blocks, an adaptive average pooling layer, a dense layer, a flattening operation, and a normalization layer. The reshape operation serves to transform the input vector $\boldsymbol{\phi}^l\in\mathbb{C}^{N\times 1},\,l\in\mathcal{L}$, into the corresponding matrix in the shape of $n_{\mathrm{max}} \times n_{\mathrm{max}}$, while the flatten operation is employed to streamline the subsequent dense layer processing. Each block is formed by the interweaving connection of two convolutional layers of a $3\times 3$ kernel size, two layer-normalization (LN) layers, and a residual connection with a convolutional layer of a $1\times 1$ kernel size. The output channels of each convolutional layer are consistent. The adaptive average pooling layer restricts the output feature map size of the block to $3\times 3$. The power allocation branch is composed of three dense layers. Finally, the outputs from the two branches are concatenated to yield an action. The critic network starts with two input branches, each comprising a dense layer. Their outputs are summed as input to alternating dense and LN layers.

The dense layer neuron quantity exceeds input and output dimensions, based on the SIM meta-atom count and data streams. The normalization layer is exploited to restrict the output of the actor neural network such that the module of meta-atoms of the metasurface layers is 1, i.e., $|\phi^l_n|^2 = 1, l \in \mathcal{L}, n \in \mathcal{N}$. The LN layer is designed to overcome the influence of the data distribution change caused by the variations in the previous layer on the neural network, which is capable of accelerating the learning speed and improving the robustness of the neural network. In contrast to the batch normalization layer that normalizes data across the batch, the LN layer performs normalization individually for each sample, maintaining independence among samples. This property endows it with superior performance in handling sequential data.

On the other hand, the \emph{LeakyReLu} activation function is utilized after the hidden dense layer and convolutional layer, while the \emph{Tanh} function is employed before the normalization layer in the branch processing the SIM phase shifts. The \emph{Softmax} function is inserted at the end of the power allocation branch for power normalization. The \emph{Adam} optimizer \cite{huang2020reconfigurable_DRL} is adopted to update the parameters of both the actor and critic training neural networks. 

Regarding the learning rate, if the model exhibits no improvement even after $\iota_p$ iteration steps and the reward ceases to increase, we multiply the learning rate by a decay factor of $\iota_f$ to reduce the learning rate. This approach enhances the model’s convergence stability toward a near-optimal solution.

\subsection{Proposed DRL-based Solution of (P1)}
In our DRL approach, an agent is assigned to continuously collect channel coefficients $\mathbf{g}_m^H,\,m\in \mathcal{M}$ and combine action $\mathbf{a}_{t-1}$ in experience replay memory to form current state information $\mathbf{s}_t$. At the very beginning of the algorithm, we need to create four neural networks, i.e., the actor training neural network $\pi(\mathbf{s}_t; \theta_\pi)$, the actor target neural network $\tilde \pi(\mathbf{s}_t; \theta_{\tilde \pi})$, critic training neural network $Q(\mathbf{s}_t, \mathbf{a}_t; \theta_q)$, and critic target neural network $\tilde Q(\mathbf{s}_t, \tilde{\mathbf{a}}_t;\theta_{\tilde q})$, and then initialize the parameters of $\pi(\mathbf{s}_t; \theta_\pi)$ and $Q(\mathbf{s}_t, \mathbf{a}_t; \theta_q)$ i.e., $\theta_q$ and $\theta_\pi$, by random sampling in a continuous space. In contrast, the parameters of $\tilde \pi(\mathbf{s}_t; \theta_{\tilde \pi})$ and $\tilde Q(\mathbf{s}_t, \tilde{\mathbf{a}}_t;\theta_{\tilde q})$ are given by (\ref{equ:target_update_c}) and (\ref{equ:target_update_a}), respectively. Moreover, the experience replay memory $M_{\mathrm{er}}$ with the capacity $C_{\mathrm{er}}$ is also established.

\begin{algorithm}[!htb]
 \caption{The DRL-Based SIM Optimization and Power Allocation}
 \label{alg}
 \begin{algorithmic}[1]
 \REQUIRE
 
 $\mu$, $\eta_c$, $\eta_a$, $\gamma_c$, $\gamma_a$, $C_{\mathrm{er}}$, $N_B$, $s_a$, $v_0$, $\zeta$ $t_{\mathrm{gap}}$, $E$, $T$.\\
 
 \ENSURE
 
 $\{ \boldsymbol{\phi}^{1}_{\mathrm{op}}, \boldsymbol{\phi}^{2}_{\mathrm{op}}, ...,\boldsymbol{\phi}^{L}_{\mathrm{op}} \}$, $\mathbf{P}_{\mathrm{op}}$, and $C_{\mathrm{op}}(\mathbf{P}, \boldsymbol{\Phi}^l)$.

 \renewcommand{\algorithmicrequire}{\textbf{Initialization:}}
 \REQUIRE

 Randomly initialize the critic training network parameter $\theta_q$ and actor training network parameter $\theta_\pi$. Initialize the critic target network parameter $\theta_{\tilde q}$ and actor target network parameter $\theta_{\tilde \pi}$ by (\ref{equ:target_update_c}) and (\ref{equ:target_update_a}), respectively. Initialize the empty experience replay memory $M_{\mathrm{er}}$.
 
 \FOR{episode $e=1,2,...,E$}
 \STATE Collect the current CSI $\{ \mathbf{g}_{1}^H, ..., \mathbf{g}_{M}^H \}$, randomly initialize the SIM coefficient matrix $\{ \boldsymbol{\phi}^{1}, \boldsymbol{\phi}^{2}, ...,\boldsymbol{\phi}^{L} \}$ and initialize transmit power matrix as $\mathbf{P}=\sqrt{\frac{P}{M}}\mathbf{I}_{M}$ to obtain initial state $\mathbf{s}_0$.
 \STATE Initialize the whitening process $\mathcal{W}$.
 
 \FOR{$t=1,2,...,T$}
 \STATE Obtain current action $\mathbf{a}_t = \pi(\mathbf{s}_t;\theta_\pi)$.
 \STATE White $\mathbf{a}_t$ by (\ref{equ:add_noise}) and then normalize it to satisfy condition (\ref{problem_2}).
 \STATE Reform $\mathbf{a}_t$ into SIM phase shifts matrix $\{ \boldsymbol{\phi}^{1}, \boldsymbol{\phi}^{2}, ...,\boldsymbol{\phi}^{L} \}$ and transmit power $\{ p_1, ..., p_M \}$. 
 \STATE Observe new state $\mathbf{s}_{t+1}$ and instant reward $r_t$ from environment given action $\mathbf{a}_t$.
 \STATE Store tuple ($\mathbf{s}_t, \mathbf{a}_t, r_t, \mathbf{s}_{t+1}$) to experience replay memory.
 \IF{the experience replay memory is full}
 \STATE Sample a $N_B$-size mini-batch tuples ($\mathbf{s}_n$, $\mathbf{a}_n$, $r_n$, $\mathbf{s}_{n+1}$) randomly from $M_{\mathrm{er}}$.
 \STATE Update the critic and actor training network, i.e., $Q(\mathbf{s}_t,\mathbf{a}_t;\theta_q)$ and $\pi(\mathbf{s}_t;\theta_\pi)$, by (\ref{equ:critic_update}) and (\ref{equ:actor_update}), respectively.
 \STATE Update the critic and actor target network, i.e., $\tilde Q(\mathbf{s}_t,\tilde{\mathbf{a}}_t;\theta_{\tilde q})$ and $\tilde \pi(\mathbf{s}_t;\theta_{\tilde \pi})$, by (\ref{equ:target_update_c}) and (\ref{equ:target_update_a}), respectively.
 \STATE Update the variance of whitening process by (\ref{equ:variance}).
 \ENDIF
 \STATE Update the state $\mathbf{s}_t$.
 \ENDFOR 
 \ENDFOR 
 \end{algorithmic} 
\end{algorithm}

The algorithm executes for $E$ episodes and iterates $T$ steps in each episode. At episode start, the agent collects CSI, resets the experience buffer and whitening noise, randomly initializes SIM phase shifts from $0$ to $2\pi$ for $\mathbf{B}$, and equally allocates transmit antenna power. During a specific episode, the agent first obtains the initial state $\mathbf{s}_0$. Then, taking the state $\mathbf{s}_t$ into the actor training network to output corresponding action $\mathbf{a}_t$. The agent obtains the reward $r_t$ by (\ref{equ:sum_rate}) and the next state $\mathbf{s}_{t+1}$ from the wireless environment. Subsequently, storing a tuple ($\mathbf{s}_t$, $\mathbf{a}_t$, $r_t$, $\mathbf{s}_{t+1}$) into experience replay memory $M_{\mathrm{er}}$. When the number of stored tuples attain the capacity $C_{\mathrm{er}}$ of the $M_{\mathrm{er}}$, signifying a full replay buffer, the critic and actor training networks start to randomly sample mini-batches of size $N_B$ from $M_{\mathrm{er}}$ and update their parameters utilizing (\ref{equ:critic_update}) and (\ref{equ:actor_update}), respectively. Ultimately, the critic target network and actor target network are updated through a soft update method, as outlined in (\ref{equ:target_update_c}) and (\ref{equ:target_update_a}), respectively. 

Finally, the optimal SIM phase shifts $\{ \boldsymbol{\phi}^{1}_{\mathrm{op}}, \boldsymbol{\phi}^{2}_{\mathrm{op}}, ...,\boldsymbol{\phi}^{L}_{\mathrm{op}} \}$ and the optimal transmit power allocation strategy $\mathbf{P}_{\mathrm{op}}$ are directly derived from the action corresponding to the maximized sum rate $C_{\mathrm{op}}(\mathbf{P}, \boldsymbol{\Phi}^l)$, which is associated with the largest instantaneous reward in the current episode. The proposed design is detailed in Algorithm 1.

\subsection{Complexity Analysis}
The computational complexity analysis of our proposed DRL algorithm can be divided into the training and predicting phases. The former involves experience replay, whitening, and four neural networks, while the latter only relates to the actor network. Denote the complexity of the adaptive pooling layer, the LN layer, and the activation function as $v_{\mathrm{pool}}$, $v_{\mathrm{ln}}$, and $v_{\mathrm{activation}}$, respectively. 

\emph{1) Training phase}: The convolutional layer complexity is encompassed within the actor network calculations. As detailed in \cite{he2015convolutional}, the complexity considering the output feature map size can be expressed as $c_{l-1} \times s_c^2 \times c_l \times s_f$, where $c_{l-1}$ denotes the number of channels in the previous layer, $s_c$ is the convolutional kernel size, $c_l$ refers to the output channel number, and $s_f$ indicates the output feature map size. Thus, the complexity of the two blocks can be expressed as $v_{\mathrm{blocks}} = 2 \times (L \times 3^2 \times c \times N) + 2 \times (c \times 3^2 \times c \times N) + 2\times (c \times 1^2 \times 2L \times N) + 6v_{\mathrm{activation}} + 4v_{\mathrm{ln}} \approx \mathcal{O}(NLc + Nc^2 + v_{\mathrm{ln}})$,
where $c$ denotes the adopted input/output channel of the convolutional layer. The adaptive pooling layer maintains the number of channels in the output feature map of the block, while decreasing the dimensions to $3\times3$ to reduce the computational complexity. The complexity of the normalization layer is proportional to $2NL$, as each meta-atom needs to calculate its phase once and then apply the modulus normalization. Moreover, according to \cite{sidelnikov2018equalization}, the overall complexity of the actor network is characterized by
\begin{equation}
 \begin{aligned}
 \label{equ:complexity_actor}
 c_{\mathrm{actor}}&=v_{\mathrm{blocks}}+v_{\mathrm{pool}}+3\times3\times c \times u_{a,0}+ \sum^3_{i=1} u_{a,i} u_{a,i+1} \\ 
 & \quad \, + 12v_{\mathrm{activation}} + 2NL\\
 &\approx \mathcal{O} \Big(NLc + Nc^2 + cu_{a,0} + \sum^3_{i=1} u_{a,i} u_{a,i+1} + v_{\mathrm{ln}} \Big),
 \end{aligned}
\end{equation}
where $u_{a,0}$ denotes the node count in the dense layer of the branch that handles phase shifts for the SIM, and $u_{a,i},\,i=\{1,2,3\}$ denotes the node count in the dense layers of the branch that allocates power in the actor network. Similarly, the complexity of the critic network can be expressed as
\begin{equation}
 \begin{aligned}
 \label{equ:complexity_critic}
 c_{\mathrm{critic}}&=\sum^3_{i=0} u_{c,i} u_{c,i+1} + 3v_{\mathrm{activation}} + 2v_{ln} \\
 &\approx \mathcal{O}\Big(\sum^3_{i=0} u_{c,i} u_{c,i+1} + v_{\mathrm{ln}} \Big),
 \end{aligned}
\end{equation}
where $u_{c,i}$ represents the node count in the $i$-th dense layer of critic network. $u_{c,0}$ equals the input size $|\mathbf{a}| + |\mathbf{s}|$, where $|\mathbf{a}|$ and $|\mathbf{s}|$ are the size of the action and state, respectively. 

On the other hand, the complexity of the whitening process is indeed related to the size of the action. Each SIM meta-atom's coefficient is whitened and then renormalized to meet the conditions in (\ref{problem_2}). Thus, the computational complexity of the whitening process is $NL + 2NL$, where $2NL$ corresponds to the cost of the renormalization operation. Therefore, the overall complexity of the training phase is
\begin{equation}
 \begin{aligned}
 \label{equ:complexity_train}
 &2 \times c_{\mathrm{critic}} + 2 \times c_{\mathrm{actor}} + NL + 2NL \\
 =&\mathcal{O} \Big (NLc + Nc^2 + cu_{a,0} + \sum^3_{i=1} u_{a,i} u_{a,i+1} \\
 & + \sum^3_{i=0} u_{c,i} u_{c,i+1} + v_{\mathrm{ln}} \Big ).
 \end{aligned}
\end{equation}

\emph{2) Predicting phase}: Since the critic network and the experience replay are adopted to ensure the actor network a faster and more stable training process, the complexity of the predicting phase only needs to consider the actor network. Therefore, the complexity of the predicting phase is
\begin{equation}
 \begin{aligned}
 \label{equ:complexity_predict}
 \mathcal{O}\Big(NLc + Nc^2 + cu_{a,0} + \sum^3_{i=1} u_{a,i} u_{a,i+1} + v_{\mathrm{ln}} \Big).
 \end{aligned}
\end{equation}

Note that the computational complexity of the DRL scheme exhibits linear scaling with the number of meta-atoms and layers in the SIM during both the training and predicting phases. 
Conversely, the traditional AO method has the complexity order of $\mathcal{O}(M^2 N^4 + MN^3L)$, and its performance heavily depends on user quantity, demonstrating nonlinear growth as system load or the number of meta-atoms increases. In comparison, the DRL method shows superior robustness to variations in both user and meta-atom quantities, offering enhanced computational efficiency and scalability, making it particularly suitable for large-scale joint parameter optimization in SIM-assisted wireless communications.
Moreover, DNN implementations in practice are highly parallelizable and run on suitable hardware (e.g., graphics processing units). Therefore, the computational complexity is practically reduced by a near-linear factor.

\section{Simulation Results and Discussion}
This section presents the numerical evaluation of the proposed DRL-optimized SIM-assisted multi-user downlink MISO wireless communication systems.
\subsection{Setup and Benchmarks}

\begin{figure}[!t]
 \centering
 \includegraphics[width=0.8\linewidth]{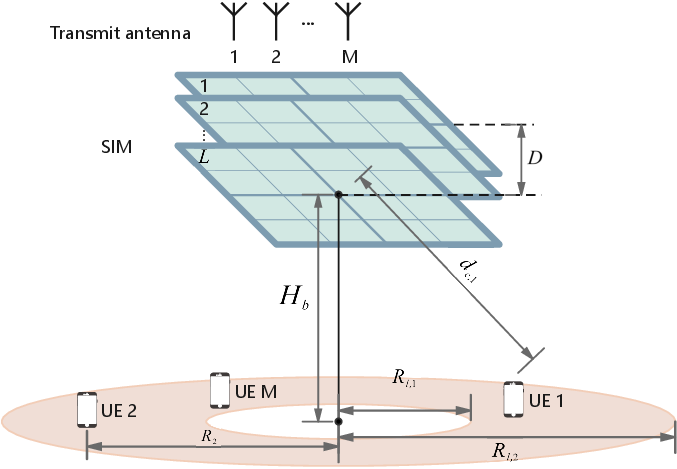}
 \caption{Simulation setup of a SIM-assisted multi-user MISO system.}
 \label{fig:location} 
\end{figure}

As shown in Fig. \ref{fig:location}, we consider a SIM-assisted multi-user MISO system that operates in the downlink at a frequency of 28 GHz, corresponding to a wavelength of $\lambda=10.7$ mm. The thickness of the SIM is set to $D = 5 \lambda$ \cite{an2023stacked_6g, an2023stacked}, consisting of $L = 4$ layers of isomorphic square metasurfaces, with each layer consists of $N = 49$ meta-atoms. The area of each meta-atom is set to $s_a=\lambda^2 / 4$. In addition, we set the distance between meta-atoms to be $r_e = \lambda / 2$. The considered simulation setup is depicted in Fig. \ref{fig:location}, where the BS is at the height $H_b=10$ meter and a SIM is integrated with the BS to facilitate wave-based precoding. The $M$ UEs are randomly distributed in an annular region as shown in Fig. \ref{fig:location} at the beginning of every episode, with the center of the SIM projected onto the ground serving as the annular center. The annulus has an inner radius $R_{l, 1}=100$ meter and an outer radius $R_{l,2}=250$ meter. Due to the random distribution of the UEs, we calculate the path losses independently using (\ref{path_loss}), where we set $C_0=-35$ dB and $\alpha=3.5$. The transmission distance of the $m$-th UE can be obtained as $d_{c,m}=(H_b^2 + R_{m}^2)^{\frac{1}{2}}$, where $R_{m}$ represents the horizontal distance from the $m$-th UE to the annular center. The Rician factor is set to $R_H=-30$ dB. At the transmitter, we set the maximum transmit power to $P=10$ dBm and the noise power is set to $\sigma_m^2=-104 \; \text{dBm},\; \forall m \in \mathcal{M}$. Additionally, we adaptively adjust the number of neurons of the dense layer based on the number of SIM meta-atoms. This strategy aims to mitigate the potential degradation in system performance due to insufficient network capacity. The other relevant parameters are presented in Table II.

\begin{table}[t]
\centering
\label{tab:parameter}
\caption{Hyper Parameters of DRL}
\begin{tabular}{lll}
\toprule 
Parameter & Meaning & Value\\
\midrule 
$E$ & The number of episodes & 50\\
$T$ & The number of iterations & 26000\\
$C_{\mathrm{er}}$ & The capacity of experience replay & 5000\\
$\zeta$ & The decay rate of whitening process & 0.95\\
$v_0$ & The initial value of whitening process & 2\\
$w_a$ & The truncation range of whitening process & 2\\
$t_{\mathrm{gap}}$ & The gap factor for discounting whitening process & 100\\
$\eta_c$ & The critic target network soft update parameter & 0.01\\
$\eta_a$ & The actor target network soft update parameter & 0.01\\
$\gamma_c$ & The critic training network initial learning rate & 0.0004\\
$\gamma_a$ & The actor training network initial learning rate & 0.0004\\
$\mu$ & The discounting factor of reward & 0.99\\
$N_B$ & The size of mini-batch & 32\\
$\iota_p$ & The patience factor for discounting the learning rate & 200 \\
$\iota_f$ & The learning rate decay factor & 0.8\\
\bottomrule 
\end{tabular}\vspace{-0.5cm}
\end{table}

To verify the performance of the proposed scheme, we consider the following four benchmark schemes for comparison:

1) \emph{DRL-UPA}: Optimize the SIM phase shifts based on the proposed DRL algorithm considering the uniform transmit power allocation.

2) \emph{Random}: The phase shifts of the SIM are configured randomly, while the transmit power of different antennas is allocated via the iterative water-filling algorithm, which follows the water-filling principle \cite{deng2009capacity} in iteration $t$ as
\begin{equation}
 \label{equ:water-filling}
 p_{m,w}^{(t)} = \left ( p_w^{(t)} - 
 \frac{\sum_{k,k \neq m}^{M} |\mathbf{g}_m^T \mathbf{b}_k |^2 p_{k,w}^{(t-1)} + \sigma^2}{
 |\mathbf{g}_m^T \mathbf{b_m} |^2} \right ), \; m \in \mathcal{M},
\end{equation}
where the power allocated to data stream $m$ in iteration $t$ is denoted as $p_{m,w}^{(t)}$ and the water level $p_w^{(t)}$ is determined such that $\sum_{m=1}^{M} p_{m,w}^{(t)} = P$.

3) \emph{Codebook}: The codebook size corresponds to the number of iterations in the proposed DRL scheme, and for each codeword, we utilize the \emph{Random} scheme. Its performance is identified as the maximum sum rate attained across all the codeword configurations explored during training. 

4) \emph{AO}: Perform iterative water-filling and gradient ascent approaches to separately optimize transmit power allocation and SIM phase shifts as in \cite{an2023stacked}.

All results are obtained through an average of 100 independent random channel realizations.

\subsection{Performance Evaluation of the Proposed Algorithm}

\begin{figure}[!t]
 \centering
		\includegraphics[width=0.9\linewidth]{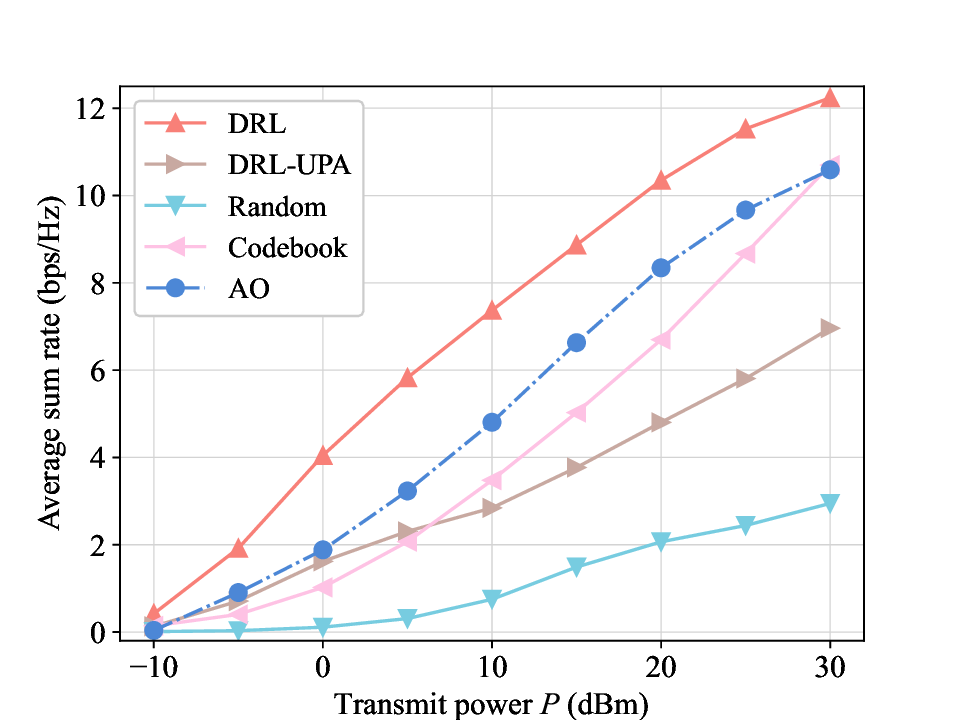}
		\caption{Average sum rate versus the transmit power $P$ for SIM-assisted schemes considering $M=4,\, L=4,\,\text{and}\, N=81$.}
		\label{fig:pt_multi_scheme}
\end{figure}
In Fig. \ref{fig:pt_multi_scheme}, we depict the average sum rate versus the transmit power $P$ at the BS, by considering $M=4$, $L=4$, and $N=81$. As observed, the joint optimization of SIM phase shifts and transmit power allocation via \emph{DRL} effectively harnesses inter-user interference, yielding superior sum rate performance, of about 2 bps/Hz, compared to the considered \emph{AO} algorithm for transmit power larger than $P=0$ dBm. Compared to \emph{DRL-UPA}, the sum rate of \emph{DRL} grows faster with the increase of transmit power, benefiting from an efficient power allocation strategy.

\begin{figure}[!t]
 \centering
		\includegraphics[width=0.9\linewidth]{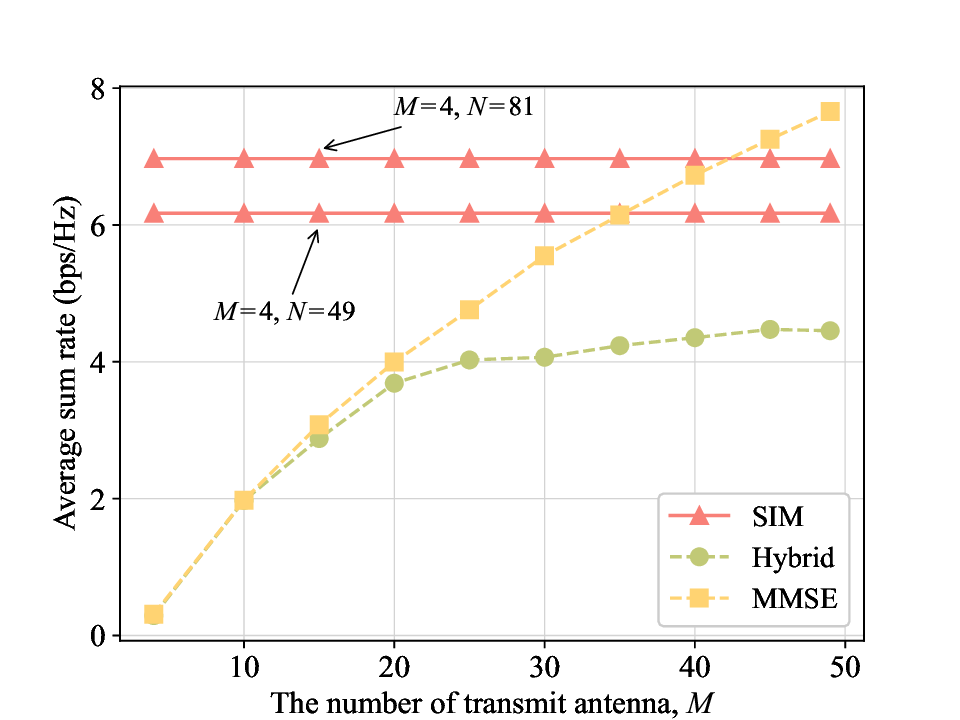}
		\caption{Average sum rate versus the number of transmit antennas for different techniques, where the number of RF chains required for the SIM and MMSE schemes are fixed to $4$, while the number of RF chains for the MMSE scheme is equal to the number of transmit antennas.}
		\label{fig:pt_diff_scheme}
\end{figure}

In Fig. \ref{fig:pt_diff_scheme}, we present the average sum rate versus the number of transmit antennas when adopting different beamforming techniques. Specifically, we consider the fully digital \emph{MMSE} precoding schemes to serve $4$ UEs, adopting water-filling transmit power allocation without employing a SIM. Additionally, we consider an advanced hybrid precoding technique \cite{zhang2018hybrid} employing $4$ RF chains to serve $4$ UEs. Notably, when considering $4$ RF chains at the transmitter, the utilization of SIM relying on DRL significantly improves the sum rate compared to the digital and hybrid precoding schemes. In this scenario, interference cancellation is entirely achieved by the SIM in the wave domain, eliminating the need for digital precoding at the BS, which enhances system performance and energy efficiency while reducing system load and hardware complexity. Furthermore, an all-digital system requires $M = 35$ transmit antennas to surpass the performance of a SIM-assisted system with $M=4$ antennas, when considering $N = 49$ low-cost meta-atoms on each layer. Moreover, with an equal number of RF chains, the SIM scheme at $N = 49$ offers greater processing flexibility and solution space than the hybrid scheme at $M = 49$, achieving a 38.5\% higher sum rate performance.

\begin{figure}[!t]
 \centering
		\includegraphics[width=0.9\linewidth]{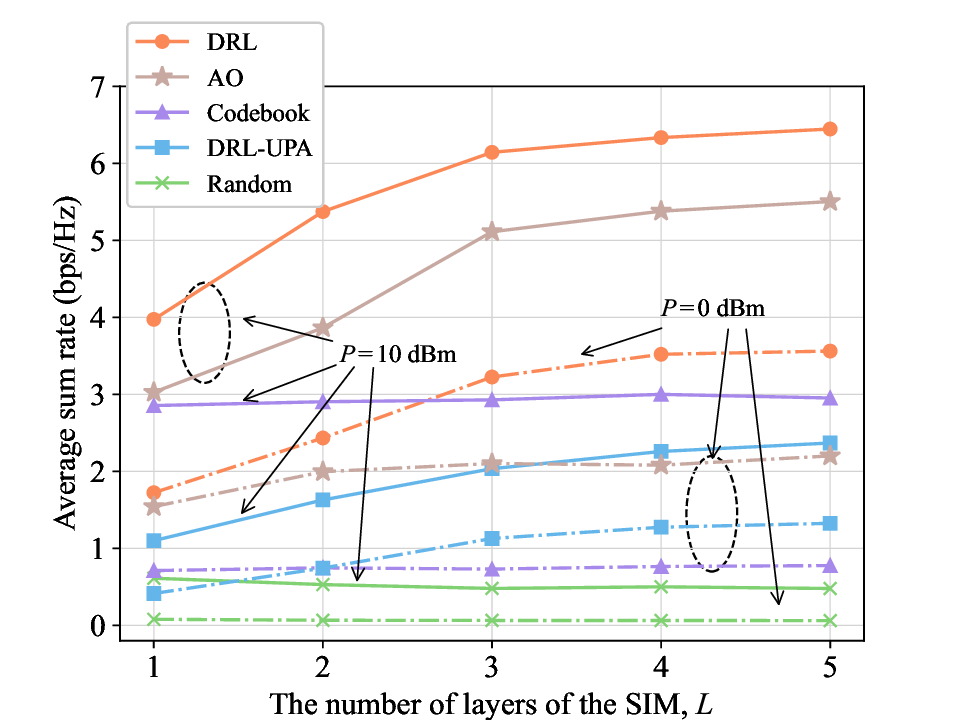}
		\caption{Average sum rate versus the number of layers $L$ considering $P=\{ 10, 0\}\, \text{dBm}$, $M=4,\text{and}\,N=49$.}
		\label{fig:layer_sum_rate}
\end{figure}

Fig. \ref{fig:layer_sum_rate} illustrates the sum rate versus the number $L$ of SIM layers for two transmit power levels. As shown, the proposed \emph{DRL} scheme consistently outperforms the considered SIM-aided benchmark schemes in both considered setups. In particular, the average sum rate initially increases and then becomes saturated when the number of SIM layers increases. The initial improvement trend stems from the enhanced interference mitigation precoding capability offered by the multi-layer structure of SIM. Then, the gains diminish and eventually become saturated, since optimizing the numerous parameters becomes intractable.

\begin{figure}[!t]
 \centering
		\includegraphics[width=0.9\linewidth]{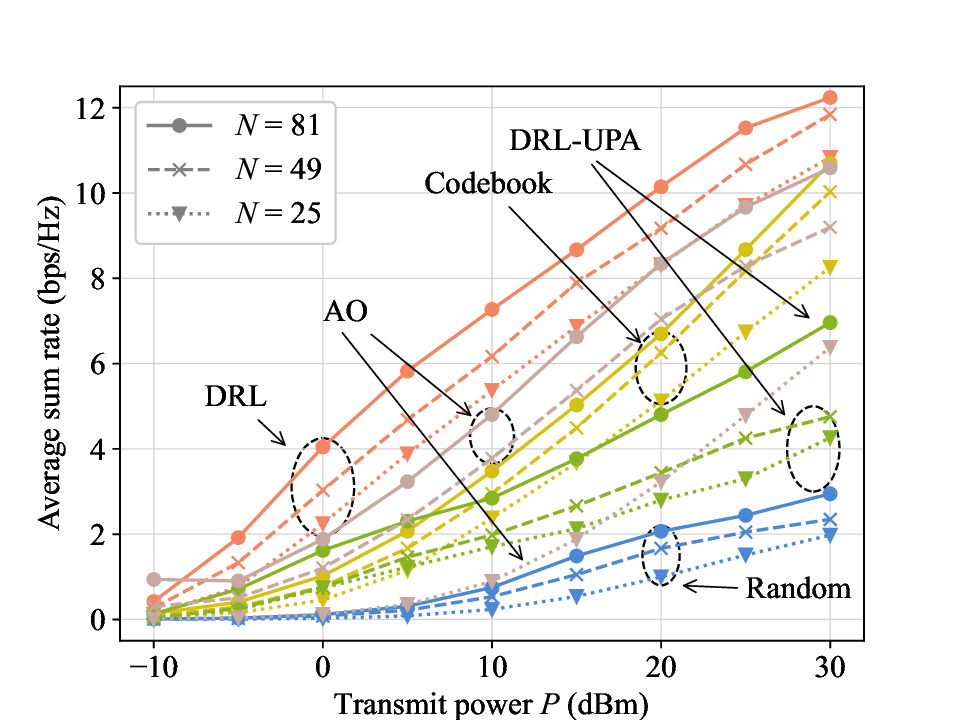}
		\caption{Average sum rate versus transmit power $P$, considering $ N=\{81,\, 49,\, 25\}$, $M=4,\text{and}\,L=4$.}
		\label{fig:pt_sum_rate}
\end{figure}

We show in Fig. \ref{fig:pt_sum_rate} the average sum rate versus the transmit power $P$ considering three system settings, i.e., $N=81$, $N=49$, and $N=25$. The proposed \emph{DRL} algorithm demonstrates superior performance compared to the other three benchmark schemes. At low transmit power levels, the proposed \emph{DRL} scheme achieves substantially higher sum rates compared to \emph{Codebook} and \emph{AO}. Specifically, the proposed scheme with $N = 25$ achieves a sum rate performance comparable to the \emph{AO} algorithm with $N = 81$, since the \emph{AO} algorithm tends to trap into some local suboptimal solutions \cite{an2023stacked}. In particular, \emph{DRL}, benefiting from DNN's powerful computational power, achieves a four-fold increase in average sum rate over \emph{Codebook} at $P=0$ dBm. However, the elevated transmit power induces an extensive dynamic range at the neural network output of \emph{DRL}, which directly degrades the performance of the scheme, diminishing the gap versus \emph{Codebook} scheme. Moreover, setting a large codebook size enables the \emph{Codebook} scheme to outperform the \emph{AO} at low $N$.

\begin{figure}[!t]
 \centering
 \includegraphics[width=0.9\linewidth]{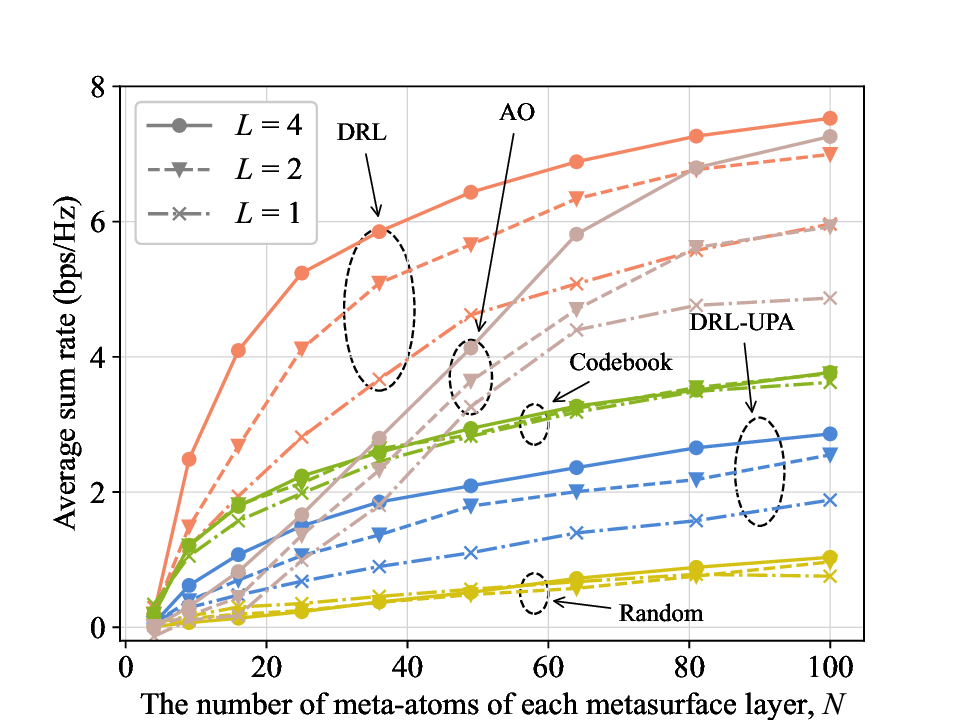}
 \caption{Average sum rate versus the number of meta-atoms of each metasurface layer, where we consider $L=\{4,\,2,\,1\}$, and $M=4$.}
 \label{fig:atom_sum_rate}
\end{figure}

In Fig. \ref{fig:atom_sum_rate}, we show the sum rate versus the number $N$ of meta-atoms per layer considering three cases for the number $L$ of SIM layers, while the other hyperparameters are identical to those described in Section IV-A. It can be seen that the sum rate increases with the number of meta-atoms per layer, showing diminishing returns beyond a threshold, e.g., $N=80$. In this regime, the spatial gain attained through the SIM approaches saturation. Specifically, the \emph{AO} algorithm gradually approaches the performance of the \emph{DRL} scheme in terms of sum rate, at the cost of an increased computational complexity, which is proportional to the number of meta-atoms. Additionally, both \emph{Random} and \emph{Codebook} with random SIM phase shifts schemes attain only mild performance gains with the increase of $L$, as they are inefficient in utilizing the degrees of freedom offered by increasing $N$.

\begin{figure}[!t]
 \centering
 \includegraphics[width=0.9\linewidth]{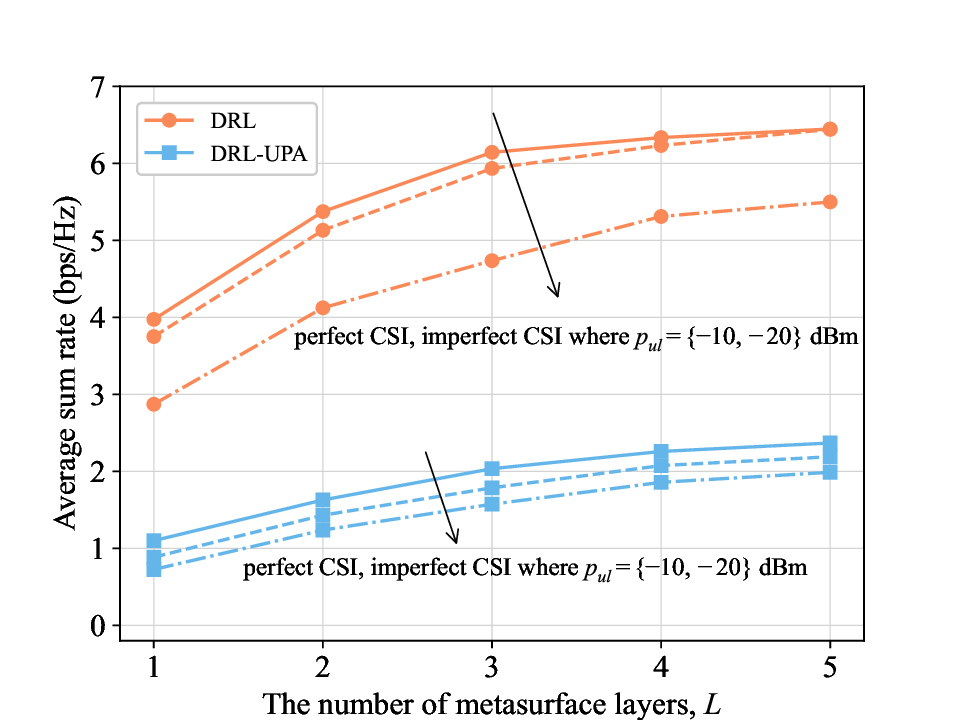}
 \caption{Average sum rate versus the number of layers $L$, where we consider imperfect CSI condition, $M=4$ and $N=49$.}
 \label{fig:imperfect_CSI}
\end{figure}
To evaluate the performance of the proposed DRL scheme under imperfect CSI, we define the uplink pilot power as $ P_{ul} $ and set the noise power to $ \sigma_{ul} = -50 $ dBm. Fig. \ref{fig:imperfect_CSI} presents the sum rate performance of the scheme under imperfect CSI, comparing different uplink pilot powers of $ P_{ul} = -10 $ dBm and $ P_{ul} = -20 $ dBm. Note that imperfect CSI leads to the performance degradation of the proposed \emph{DRL} scheme for a small number of layers. However, this deficit diminishes progressively as the increased number of layers enhances the computational capacity. Despite achieving a relatively low sum rate, \emph{DRL-UPA} remains susceptible to CSI errors, leading to a 12\% performance degradation at $p_{ul}=-10$ dBm.

\subsection{Impact of Hyper-Parameters of DRL}

\begin{figure}[!t]
\centering
\subfloat[]{\includegraphics[width=4.6cm]{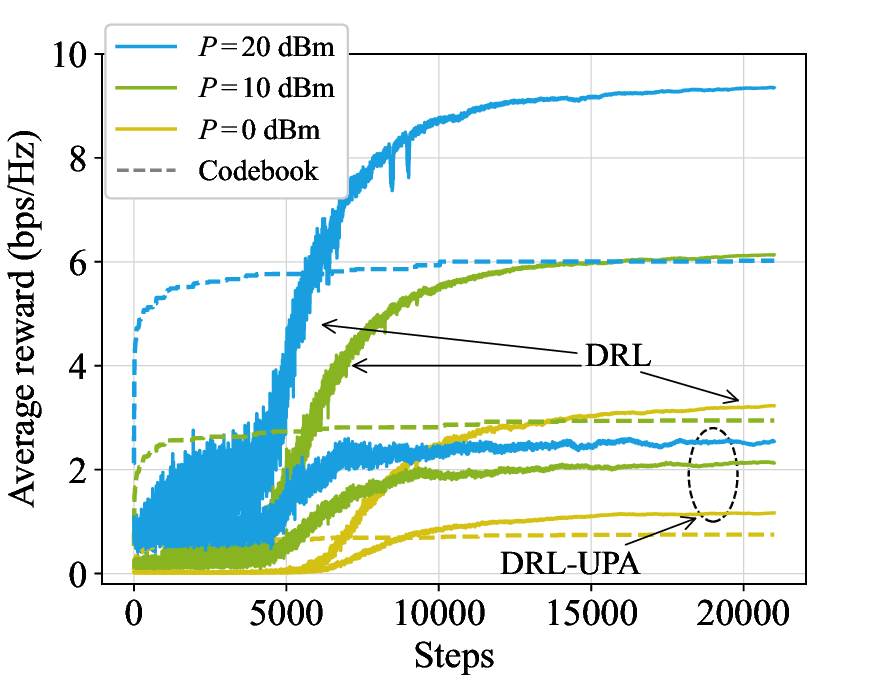}\label{fig:simulation_Pt}} 
\subfloat[]{\includegraphics[width=4.6cm]{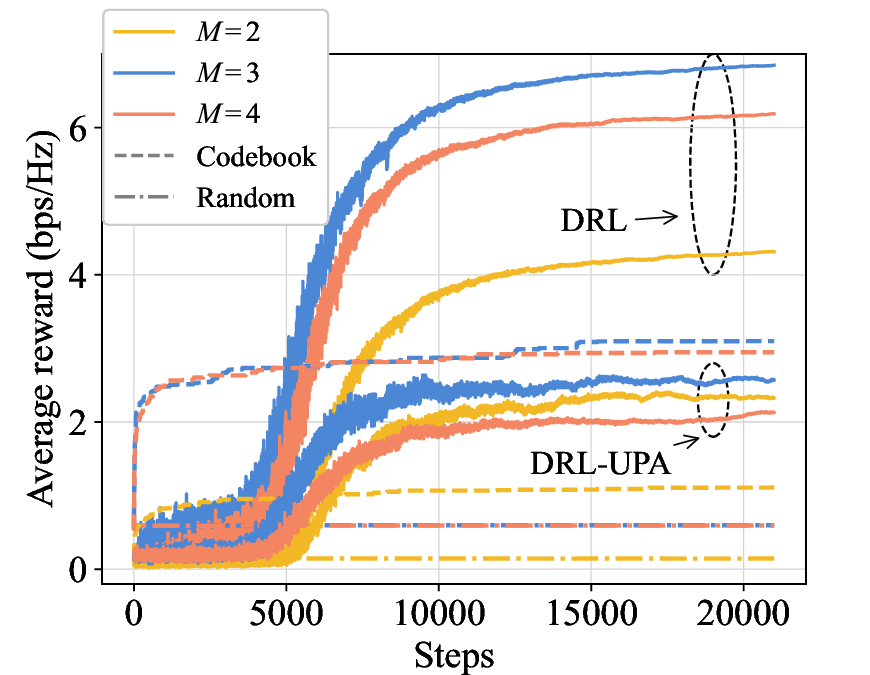}\label{fig:data_stream} }
\caption{The convergence behavior of the average sum rate, where we consider (a) $P=\{0, 10, 20\}\, \text{dBm}$, and $M=4$. (b) $M=\{2,3,4 \}$, and $P=10$ dBm.}
\end{figure}

To better demonstrate the optimization capability of the proposed \emph{DRL} algorithm, Fig. \ref{fig:simulation_Pt} shows its convergence process for three transmit power levels. It can be seen that the transmit power levels significantly affect the final convergence result. Specifically, in our proposed scheme, the improvements in average reward are considerable as the transmit power increases. This is because the proposed \emph{DRL} can jointly adapt both the SIM phase shifts and transmit power allocation strategy, thus improving the system sum rate performance. However, \emph{DRL-UPA}, which only optimizes the SIM phase shifts, has a relatively small gain from the increase of transmit power. Furthermore, we compare the average reward versus training steps in Fig. \ref{fig:data_stream} under different numbers of data streams, i.e., $M=\{2,3,4\}$, while other parameters remain the same as those described in Section IV-A. Observed from Fig. \ref{fig:data_stream} that \emph{Codebook} and \emph{Random} are unable to efficiently optimize such a massive number of parameters as the number of data streams increases. Thus, adjusting the number of data streams has less impact on \emph{Codebook} and \emph{Random} in terms of the convergence speed and resultant performance.

\begin{figure}[!t]
\centering
\subfloat[]{\includegraphics[width=4.6cm]{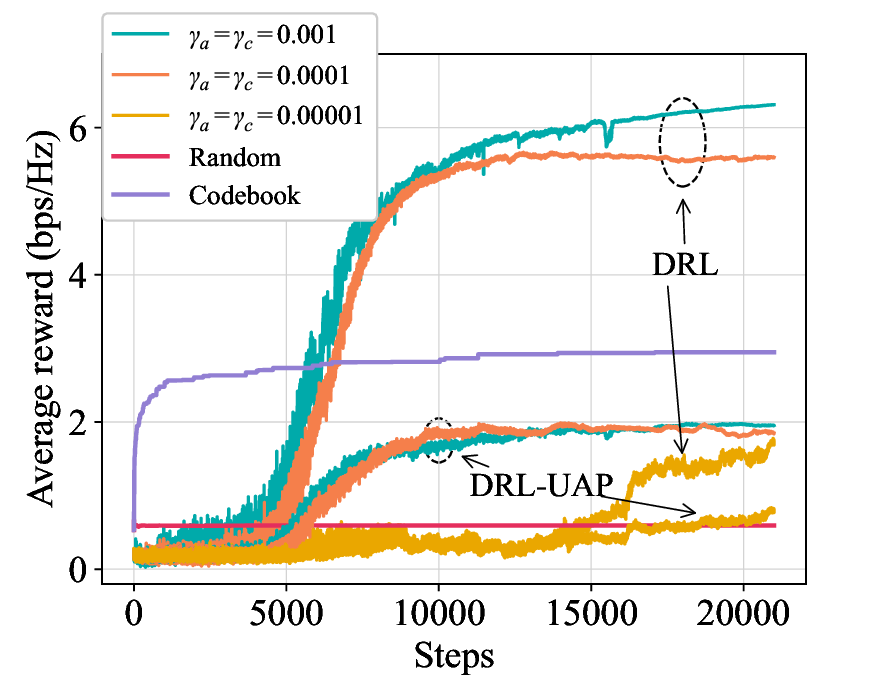}\label{fig:learning_rate}} 
\subfloat[]{\includegraphics[width=4.6cm]{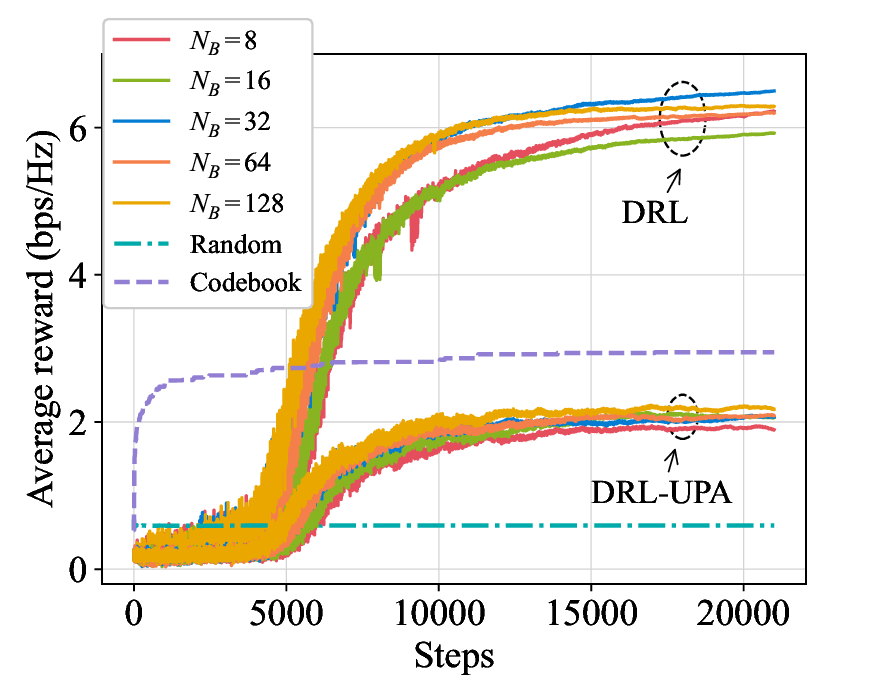}\label{fig:batch_size} }
\caption{The convergence behavior of the average sum rate, where we consider (a) $\gamma_a=\gamma_c=\{ 0.001, 0.0001, 0.00001 \}$, and $M=4$. (b) $N_B=\{ 8,16,32,64,128 \}$.}
\end{figure}

Fig. \ref{fig:learning_rate} demonstrates the effect of different learning rates, considering $\gamma_a = \gamma_c= \{0.001, 0.0001, 0.00001 \}$, and the data stream is set to $M=4$. It is demonstrated that the learning rate dictates the system's optimal performance and ability to converge. Specifically, when adopting a learning rate of 0.0001, DRL can converge to a relatively satisfactory reward level and further decrease the learning rate over the training process exhibiting a trend of continued reward optimization. This is because, after sufficient training, a relatively lower learning rate could effectively constrain the parameter space and avoid the model learning noise with a higher probability. These benefits allow the model to fine-tune the parameters to find a better solution near the local minimum. However, with a lower learning rate, such as 0.00001, the system is more likely to become trapped in a local optimum because of the lack of exploration competence. Therefore, it is crucial to determine an appropriate learning rate to ensure that the system achieves optimal performance. Moreover, the codebook scheme has a fast convergence faster but performs poorly, whereas the DRL approach achieves a superior sum rate by relying on a well-configured SIM.

Next, we show in Fig. \ref{fig:batch_size} the effect of different batch sizes considering $N_B=\{8, 16, 32, 64, 128\}$, while the transmit power is $P = 10$ dBm and the number of the data streams is $M = 4$. During the \emph{DRL} training process, the batch size does not significantly influence the resultant sum rate performance. Instead, it slightly affects the convergence speed and efficacy of the \emph{DRL} algorithm. For example, when employing a batch size of 32, the \emph{DRL} algorithm rapidly converges to the maximum. A larger batch size reduces the gradient's variance while improving the gradient's stability. In contrast, when the batch sizes are set to 8 or 16, the algorithm exhibits slower convergence behavior. The convergence behaviors of \emph{DRL-UPA} are less sensitive to the batch size, which, however, suffers from severe performance loss due to the CSI-unaware power allocation. 


\begin{table}[!t]
\centering
\caption{Ablation study of state design}
\begin{tabular}{c|c|c}
\hline
State design & \makecell[l]{Average sum rate} & \makecell[l]{Training stability} \\ \hline
$\left\{\textrm{CSI}\left ( t \right )\right\}$ & $4.62$ bps/Hz & $0.78$ bps/Hz \\ \hline
$\left\{ \textrm{CSI}\left ( t \right ), \mathbf{a}_{t-1} \right\}$ & $5.12$ bps/Hz & $0.31$ bps/Hz \\ \hline
$\left\{ \textrm{CSI}\left ( t \right ), \mathbf{a}_{t-1}, r_{t-1} \right\}$ & $5.13$ bps/Hz & $0.24$ bps/Hz \\ \hline 
\end{tabular}\label{tab:ablation}
\end{table}

To validate our proposed state design, we have performed an ablation experiment by comparing three different state configurations. The numerical results are shown in Tab. \ref{tab:ablation}, where we set $L=4$ metasurface layers, $N=49$ meta-atoms, $M=4$ data streams, and $P=10$ dBm. The average sum rate and training stability are characterized by the mean and standard deviation of the sum rate after achieving convergence, respectively. The results suggest that including historical information ($\mathbf{a}_{t-1}$, $r_{t-1}$) empirically improves policy robustness against fast fading, at the cost of the increased dimensionality of the state. Therefore, our proposed scheme adopts a state design that incorporates previous action and reward, which effectively enhances the training performance and stability, enabling the agent to learn the optimal SIM phase shift configuration and power allocation strategy in long-term trends, ultimately improving the system's sum rate.

\subsection{Impact of Whitening Process}
\begin{figure}[t]
 \centering
 \includegraphics[width=0.9\linewidth]{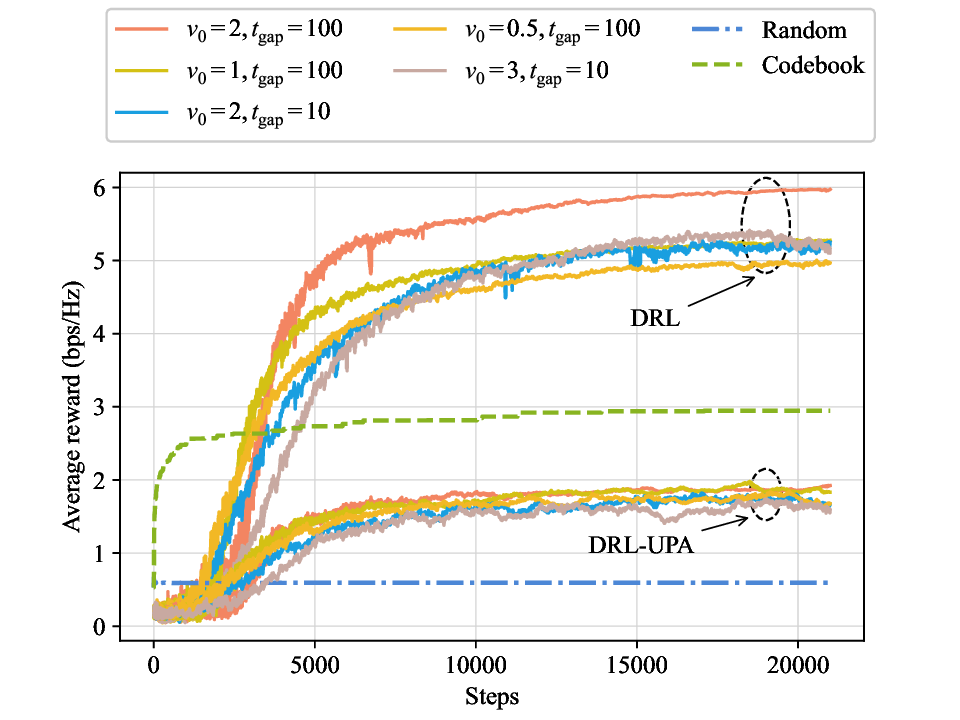}
 \caption{The convergence behavior of the average sum rate under different parameters of the whitening processes.}
 \label{fig:whitening}
\end{figure}

Finally, in Fig. \ref{fig:whitening}, we investigate the impact of the whitening process in (\ref{equ:add_noise}) on the convergence behavior of the DRL algorithm. We note that different parameters involved in the whitening process significantly influence the convergence effect of rewards. Setting $v_0=2$ and $t_{\mathrm{gap}}=100$ provides sufficient exploration ability in both early and late training stages, enabling a broader range of attempted actions circumventing over-dependence on proven high-return tactics. This parameterization led to a $20\%$ performance increase compared to the suboptimal values of $v_0=0.5, t_{\mathrm{gap}}=100$.

Additionally, the whitening process leads to a smoother \emph{DRL} training process as it simulates the uncertainty in a real wireless communications environment, thus granting a certain level of robustness to the trained model against uncharted interference. Indeed, the whitening process has the effect of regularization to reduce overfitting and improve the generalization ability of the model. Nevertheless, introducing excessive noise also impedes convergence, which requires careful controlling of the noise variance. Accordingly, attenuation of the whitening process over time is implemented, with Fig. \ref{fig:whitening} illustrating the efficacy of this attenuation strategy. According to the above analysis, the whitening process plays a pivotal role in the proposed DRL algorithm.

\section{Conclusions}
This paper investigated a SIM-assisted multi-user MISO wireless communication system profiting from the SIM-enabled precoding in the wave domain. A joint optimization framework of the SIM phase shifts and transmit power allocation, aiming to maximize the sum-rate performance, was formulated. To address this challenging non-convex optimization, a DRL approach operating on continuous value solutions and without prerequisite labeled data was proposed. The SIM phase shifts and power allocation strategies were directly extracted from the DRL’s actor network. The presented simulation results validated the efficiency of the SIM for multi-user interference suppression, showing a 38.5\% sum rate improvement over advanced hybrid precoding methods. Furthermore, the proposed DRL optimization for an indicative SIM-assisted multi-user MISO system demonstrated a 2 bps/Hz sum-rate improvement compared to a state-of-the-art AO algorithm. It was also showcased that integrating appropriate hyperparameter selection and a whitening process can substantially enhance the robustness of the proposed DRL algorithm. For future research, SIM-enabled broadband communication and discrete space optimization techniques deserve further exploration.

\bibliographystyle{IEEEtran}
\bibliography{mylib}

\end{document}